\documentclass[manuscript]{aastex}

\def\sgn{\mathop{\rm sgn}\nolimits} 

\shorttitle{Titan's interior from its obliquity}
\shortauthors{Noyelles \& Nimmo}

\usepackage{esvect}
\usepackage{lscape}
\usepackage{amsmath}
\usepackage{wasysym}

\usepackage{color}
\usepackage{graphicx}

\begin{document}


\title{New insights on Titan's interior from its obliquity}


\author{Beno\^it Noyelles}
\affil{NAmur Center for CompleX SYStems (naXys) -- University of Namur -- Rempart de la Vierge 8 -- B-5000 Namur -- Belgium}
\email{benoit.noyelles@unamur.be}

\and

\author{Francis Nimmo}
\affil{Department of Earth and Planetary Sciences -- University of California at Santa Cruz -- 1156 High Street -- Santa Cruz, California 95064 -- USA}
\email{fnimmo@es.ucsc.edu}

%
%



\begin{abstract}


\par We constructed a 6-degrees of freedom rotational model of Titan as a 3-layer body consisting of a rigid core, a fluid global ocean, and a floating ice shell. 
The ice shell exhibits partially-compensated lateral thickness variations in order to simultaneously match the observed degree-two gravity and shape coefficients. The rotational
dynamics are affected by the gravitational torque of Saturn, the gravitational coupling between the inner core and the shell, and the pressure coupling at the fluid-solid
boundaries. Between $10$ and $13\%$ of our model Titans have an obliquity (due to a resonance with the $29.5$-year periodic annual forcing) that is consistent with the observed value.

\par The shells of the successful models have a mean thickness of $130$ to $140$ km, and an ocean of $\approx$250 km thickness. Our simulations of the obliquity evolution 
show that the {\em Cassini} obliquity measurement is an instantaneous one, and does not represent a mean value. Future measurements of the time derivative of the obliquity would help to
refine the interior models. We expect in particular a variation of roughly 7 arcmin over the duration of the {\em Cassini} mission.

\end{abstract}





\section{Introduction\label{sec:intro}}

\par The {\em Cassini} spacecraft, in orbit around Saturn since July 2004, has allowed huge progress on modelling of the internal structure and the rotational dynamics of Titan. 
An internal ocean is consistent with the measurements of the tidal Love number $k_2\approx 0.6$ \citep{ijdslaarrt2012} and was theoretically predicted by 
\citet{ls1987}, this prediction being supported by several following studies, e.g. \citep{gs1996,gsd2000,tglms2005,fgtv2007}. A comparison between the shape of
Titan \citep{zshlkl2009} (Tab.\ref{tab:titanshape}) and its gravity field \citep{irjrstaa2010} (Tab.\ref{tab:titangravity}) suggests either variations in the 
thickness of a floating ice shell \citep{nb2010,hnzi2013} or lateral variations in the shell's density \citep{cs2012}.

\begin{table}[ht]
\centering
\caption[The shape of Titan.]{The shape of Titan, from \citep{zshlkl2009}.\label{tab:titanshape}}
\begin{tabular}{lr}
\hline
Parameter & Value \\
\hline
Subplanetary equatorial radius $a$ & $2575.15\pm0.02$ km \\
Along orbit equatorial radius $b$  & $2574.78\pm0.06$ km \\
Polar radius $c$                   & $2574.47\pm0.06$ km \\
Mean radius $R$                    & $2574.73\pm0.09$ km \\
\hline
\end{tabular}
\end{table}

\placetable{tab:titanshape}

\begin{table}[ht]
\centering
\caption[The 2 solutions for the gravity field of Titan.]{The 2 solutions for the gravity field of Titan \citep{irjrstaa2010}. SOL1 is a single multiarc solution 
obtained from 4 flybys of {\em Cassini} dedicated to the determination of the gravity field, while SOL2 is a more general
approach, in which all available radiometric tracking and optical navigation imaging data from the Pioneer and
Voyager Saturn encounters and astronomical observations of Saturn and its satellites are considered. The
uncertainties correspond to $1\sigma$. The global solution SOL2 could be consistent with the hydrostatic equilibrium 
($J_2/C_{22}\approx10/3$), but the shape is not \citep{zshlkl2009}. In our study of a triaxial Titan, the only coefficients 
we use are $\mathcal{G}M_6$, $J_2$ and $C_{22}$.\label{tab:titangravity}}
\begin{tabular}{lrr}
\hline
                  & SOL1                             & SOL2 \\
\hline
$\mathcal{G}M_6$  &  --                              & $8978.1394$ $km^3.s^{-2}$ \\
$J_2$             & $(3.1808\pm0.0404)\times10^{-5}$ & $(3.3462\pm0.0632)\times10^{-5}$ \\
$C_{21}$          & $(3.38\pm3.50)\times10^{-7}$     & $(4.8\pm11.5)\times10^{-8}$ \\
$S_{21}$          & $(-3.52\pm4.38)\times10^{-7}$    & $(6.20\pm4.96)\times10^{-7}$ \\
$C_{22}$          & $(9.983\pm0.039)\times10^{-6}$   & $(1.0022\pm0.0071)\times10^{-5}$ \\
$S_{22}$          & $(2.17\pm0.41)\times10^{-7}$     & $(2.56\pm0.72)\times10^{-7}$ \\
$J_3$             & $(-1.879\pm1.019)\times10^{-6}$  & $(-7.4\pm105.1)\times10^{-8}$ \\
$C_{31}$          & $(1.058\pm0.260)\times10^{-6}$   & $(1.805\pm0.297)\times10^{-6}$ \\
$S_{31}$          & $(5.09\pm2.02)\times10^{-7}$     & $(2.83\pm3.54)\times10^{-7}$ \\
$C_{32}$          & $(3.64\pm1.13)\times10^{-7}$     & $(1.36\pm1.58)\times10^{-7}$ \\
$S_{32}$          & $(3.47\pm0.80)\times10^{-7}$     & $(1.59\pm1.05)\times10^{-7}$ \\
$C_{33}$          & $(-1.99\pm0.09)\times10^{-7}$    & $(-1.85\pm0.12)\times10^{-7}$ \\
$S_{33}$          & $(-1.71\pm0.15)\times10^{-7}$    & $(-1.49\pm0.16)\times10^{-7}$ \\
$J_2/C_{22}$      & $3.186\pm0.042$                  & $3.339\pm0.067$ \\
\hline
\end{tabular}
\end{table}

\placetable{tab:titangravity}

\par {\em Cassini} observed Titan's rotation as well. The most recent measurements suggest the expected synchronous rotation \citep{mi2012} and a pretty high obliquity
of $\approx0.3^{\circ}$ at the mean date March $11^{th}$ 2007, already detected by \citep{sklhloacgiphjw2008}. If we assume that the rotation of Titan has reached its most probable dynamical 
equilibrium state, i.e. Cassini State 1, then this obliquity is not consistent with a rigid Titan \citep{nlv2008,bn2008,bn2011}.
However, the presence of an internal ocean can lead to a resonant process raising the obliquity of Titan \citep{bvyk2011}, making the high obliquity a possible signature of a
global subsurface ocean.

\par In this paper, we simulate the rotation of Titan, considering both the internal structure and all the dynamical degrees of freedom. Our Titan is a 3-layer body 
composed of a rigid inner core, a global ocean and rigid shell with a variable thickness. For each of the 2 rigid layers, we 
simulate at the same time the longitudinal motion, the orientation of the angular momentum, and of the figure polar axis. The dynamics of these 2 layers will 
be affected by the gravitational pull of Saturn, the pressure coupling at the interface with the ocean and the gravitational coupling between them. The pressure
coupling is modelled after \citet{bvyk2011} and the gravitational coupling after \citet{sx1997}. In calculating the torques, we take into account 
variations in the thickness of the ice shell \citep{nb2010} consistent with the gravity and topography constraints. We then identify interior structures for which the 
predicted rotation state is consistent with the observations, before simulating the expected behavior of the obliquity of Titan.

Our model confirms the conclusion of \citet{bvyk2011} that the unexpectedly high obliquity of Titan could be due to a resonance with the periodic 
annual forcing. We go further, however, in showing that the obliquity is predicted to be time-variable (Fig~\ref{fig:leftrightcassini}): a prediction which analysis of 
{\em Cassini} radar observations \citep{bsk2013} should be able to test.


\section{The equations of the problem}

The approach that we follow below is a generalization of the scheme adopted by \citet{bvyk2011}. There are two 
important innovations in our approach. First, we consider the three-dimensional orientation of the shell and of the core, so that we can simultaneously 
treat both obliquity \citep{bn2011,bvyk2011} and also longitudinal librations \citep{vbt2013,rrc2014}, as well as determining the 
magnitude of the usually neglected polar motion. Second, we explicitly take into 
account the rigidity and spatial variations in thickness of the ice shell, which are indicated by Titan's topography and 
gravity \citep{nb2010,hnzi2013} and which affect the resulting torques.

Although our model is quite complicated, it does neglect some potentially significant effects. Most notably, we do not 
consider the effects of either atmospheric torques or torques due to flow in the subsurface ocean. The low viscosity of water argues 
against the latter being important, but in some cases tidal forcing can lead to strong flows \citep{nhwba2009,clml2012}. We defer 
consideration of this topic to future work. As discussed below, we also neglect the potential effect of restoring torques 
due to elastic deformation of the ice shell \citep{gm2010,rrc2014}.

\subsection{Parameterization of the problem}

\par
\begin{figure}[ht]
	\centering
	\includegraphics[width=0.8\textwidth]{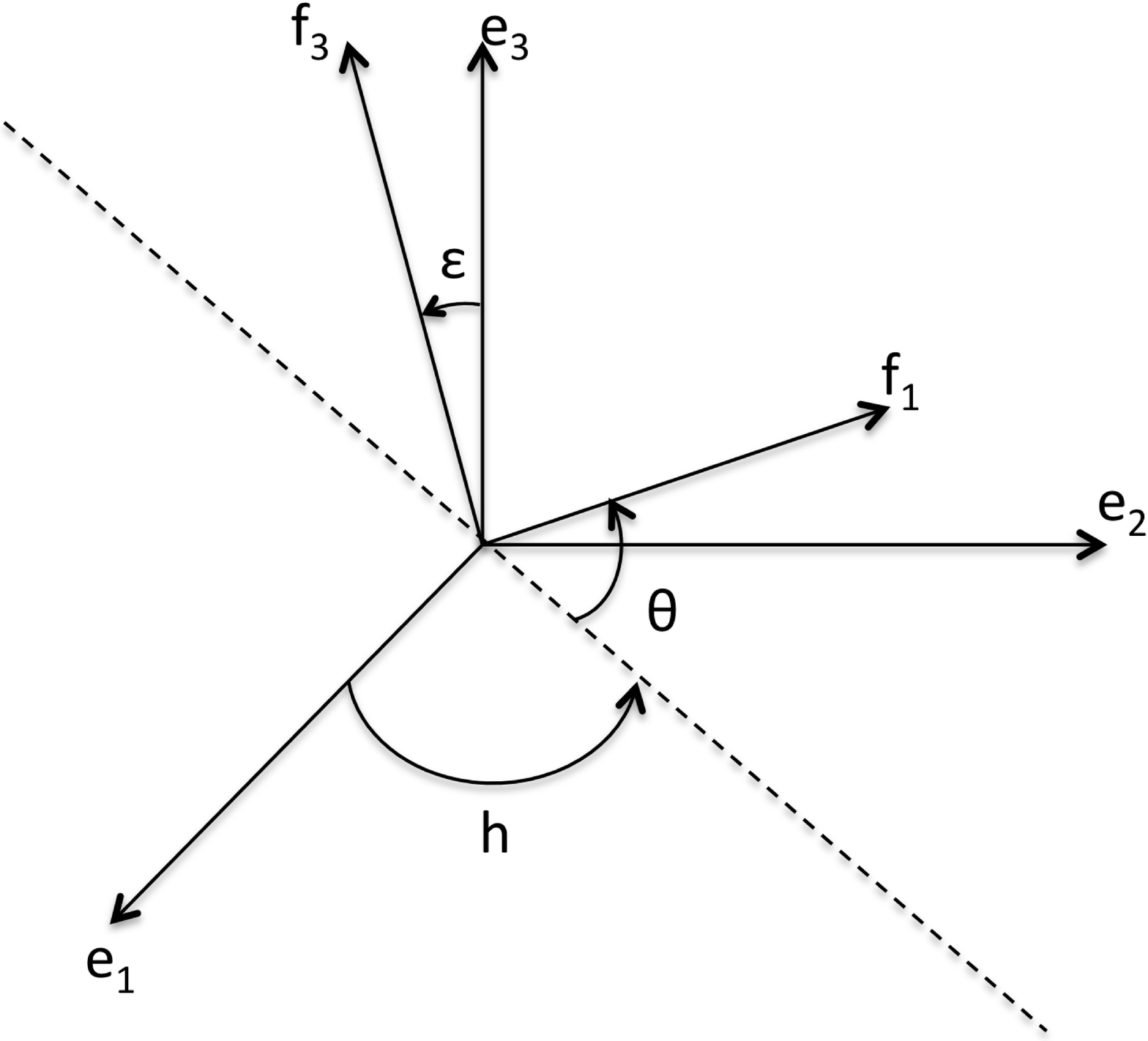}
	\caption[The Euler angles.]{Euler angles $(h,\epsilon,\theta)$ giving the orientation of the principal axes of inertia $(\vec{f_1},\vec{f_2},\vec{f_3})$ of a 
		rigid layer (inner core or shell) with respect to the inertial reference frame $(\vec{e_1},\vec{e_2},\vec{e_3})$. These quantities 
	should be written with a superscript c or s, i.e. either $h^c$ or $h^s$ whether they are related to the inner core or to the shell. 
	The dashed line is the intersection between the inertial reference plane $(\vec{e_1},\vec{e_2})$ and the equatorial plane of the core or shell $(\vec{f_1},\vec{f_2})$. \label{fig:euler}}
\end{figure}

 We simulate the orientation of both the rigid inner core and the rigid shell (or crust). For that, we need 2 sets of Euler angles, respectively 
$(h^c,\epsilon^c,\theta^c)$ for the core and $(h^s,\epsilon^s,\theta^s)$ for the shell to represent the orientation of the principal axes of inertia of the 
considered layer in an inertial reference frame (Fig.\ref{fig:euler}).

\placefigure{fig:euler}

\par These Euler angles present a virtual singularity. If the quantity $\epsilon$ is null, then the angles $h$ and $\theta$ are not uniquely defined, 
but their sum is. In practice, it appears that when $\epsilon$ is small enough, then numerical uncertainties can erronously suggest an erratic behavior. We 
by-passed this problem by using the following cartesian-like coordinates:

\begin{eqnarray}
	\xi^{s,c}  & = & \epsilon^{s,c}\sin h^{s,c}, \nonumber \\
	\eta^{s,c} & = & \epsilon^{s,c}\cos h^{s,c}, \nonumber \\
	p^{s,c}    & = & h^{s,c}+\theta^{s,c}.	\nonumber
\end{eqnarray}
When $\epsilon$ is null, then $\eta$ and $\xi$ are both null and the system does not present any singularity.

\par We also need to represent the angular momentum $\vec{G}$ of each of these rigid layers. For that we use as variables the components of the associated rotation vector 
$\vec{\omega}$; this yields
$\vec{G^s}=A^s\omega_1^s\vec{f_1^s}+B^s\omega_2^s\vec{f_2^s}+C^s\omega_3^s\vec{f_3^s}$ for the shell and 
$\vec{G^c}=A^c\omega_1^c\vec{f_1^c}+B^c\omega_2^c\vec{f_2^c}+C^c\omega_3^s\vec{f_3^c}$ for the core. $A$, $B$ and $C$ are the principal moments of inertia
of the layer under consideration; we have for the core:

\begin{eqnarray}
	A^c & = & \iiint_{core}\rho_c(x,y,z) \left(y^2+z^2\right) dxdydz, \label{eq:Ac} \\
	B^c & = & \iiint_{core}\rho_c(x,y,z) \left(x^2+z^2\right) dxdydz, \label{eq:Bc} \\
	C^c & = & \iiint_{core}\rho_c(x,y,z) \left(x^2+y^2\right) dxdydz, \label{eq:Cc}
\end{eqnarray}
and similar formulae for the shell. $\rho_c(x,y,z)$ is the density of the core, $(x,y,z)$ being the classical writings for the cartesian coordinates, in the
reference frame of the principal axes of inertia $(\vec{f_1^c},\vec{f_2^c},\vec{f_3^c})$.
	
\subsection{Kinematic equations controlling the Euler angles}

\par To determine the relations linking the Euler angles of a layer to the components of the angular momentum, we have to keep in mind the geometry of the 
problem (Fig.\ref{fig:euler}). As explained for instance in \citep{fc1999}, the rotation vector 
$\vec{\omega}=\omega_1\vec{f_1}+\omega_2\vec{f_2}+\omega_3\vec{f_3}$ represents 3 successive rotations:

\begin{enumerate}
	
\item a rotation around the z-axis (here $\vec{e_3}$) of an angle $h$,
	
\item then a rotation around the new, but not final, x-axis of angle $\epsilon$, 
	
\item and finally a rotation around the final z-axis, i.e. $\vec{f_3}$, of an angle $\theta$.
	
\end{enumerate}
This reads mathematically:

\begin{equation}
	\label{eq:kinetic1}
	\left(\begin{array}{c}
	\omega_1 \\
	\omega_2 \\
	\omega_3 
	\end{array}\right) = 
	\left(\begin{array}{c}
	0 \\
	0 \\
	\dot{\theta} 
	\end{array}\right)+R_3(-\theta)\left(\begin{array}{c}
	\dot{\epsilon} \\
	0 \\
	0
	\end{array}\right)+R_3(-\theta)R_1(-\epsilon)\left(\begin{array}{c}
	0 \\
	0 \\
	\dot{h} 
	\end{array}\right),
\end{equation}
with 

\begin{equation}
R_3(\phi)=\left(\begin{array}{ccc}
\cos\phi & -\sin\phi & 0 \\
\sin\phi & \cos\phi & 0 \\
0 & 0 & 1
\end{array}\right)
\label{equ:r3}
\end{equation}
and

\begin{equation}
R_1(\phi)=\left(\begin{array}{ccc}
1 & 0 & 0 \\
0 & \cos\phi & -\sin\phi \\
0 & \sin\phi & \cos\phi 
\end{array}\right).
\label{equ:r1}
\end{equation}

This yields 

\begin{eqnarray}
	\omega_1 & = & \dot{\epsilon}\cos\theta+\dot{h}\sin\epsilon\sin\theta, \label{eq:omega1} \\
	\omega_2 & = & -\dot{\epsilon}\sin\theta+\dot{h}\sin\epsilon\cos\theta, \label{eq:omega2} \\
	\omega_3 & = & \dot{\theta}+\dot{h}\cos\epsilon, \label{eq:omega3}
\end{eqnarray}
and

\begin{eqnarray}
	\dot{h} & = & \frac{\omega_1\sin\theta+\omega_2\cos\theta}{\sin\epsilon}, \label{eq:doth} \\
	\dot{\epsilon} & = & \omega_1\cos\theta-\omega_2\sin\theta, \label{eq:dotepsilon} \\
	\dot{\theta} & = & \omega_3-\frac{\omega_1\sin\theta+\omega_2\cos\theta}{\tan\epsilon}. \label{eq:dottheta}
\end{eqnarray}

The equations (\ref{eq:doth}) and (\ref{eq:dottheta}) illustrate the virtual singularity we mentioned above. These formulae are the same as the ones present
in \citep{bnt1995,htn2011} but are given with a different sign in numerous other studies, e.g. \citep{wbyrd2001}, probably because of a different sign 
convention. We now get straightforwardly:

\begin{eqnarray}
	\dot{\xi}  & = & \left(\omega_1\cos\theta-\omega_2\sin\theta\right)\sin h+\frac{\epsilon}{\sin\epsilon}\left(\omega_1\sin\theta+\omega_2\cos\theta\right)\cos h, \label{eq:dotxi} \\
	\dot{\eta} & = & \left(\omega_1\cos\theta-\omega_2\sin\theta\right)\cos h-\frac{\epsilon}{\sin\epsilon}\left(\omega_1\sin\theta+\omega_2\cos\theta\right)\sin h, \label{eq:doteta} \\
	\dot{p}    & = & \omega_3+\left(\omega_1\sin\theta+\omega_2\cos\theta\right)\tan\frac{\epsilon}{2}. \label{eq:dotp}
\end{eqnarray}
The virtual singularity has nearly disappeared. The only numerical problem that could remain is due to $\epsilon/\sin\epsilon$ in the Eq.(\ref{eq:dotxi})
and (\ref{eq:doteta}), so we replace it by its Taylor expansion $1+\epsilon^2/6+7\epsilon^4/360$ for $\|\epsilon\|<10^{-8}$.

\par As shown in \citep{bvyk2011}, the dynamical equations reduce to

\begin{eqnarray}
	\frac{d\vv{G^c}}{dt} & = & \vv{\Gamma^c_{\saturn}}+\vv{\Gamma^c_o}+\vv{\Gamma^c_{sh}}, \label{eq:dGc} \\
	\frac{d\vv{G^s}}{dt} & = & \vv{\Gamma^s_{\saturn}}+\vv{\Gamma^s_o}+\vv{\Gamma^s_{co}}, \label{eq:dGs}
\end{eqnarray}
the relevant torques being:

\begin{itemize}
\item $\vv{\Gamma^c_{\saturn}}$: gravitational torque of Saturn on the core,
\item $\vv{\Gamma^c_o}$: pressure coupling of the ocean at the core-ocean boundary,
\item $\vv{\Gamma^c_{sh}}$: gravitational torque of the shell on the core,
\item $\vv{\Gamma^s_{\saturn}}$: gravitational torque of Saturn on the shell,
\item $\vv{\Gamma^s_o}$: pressure coupling of the ocean at the shell-ocean boundary,
\item $\vv{\Gamma^s_{co}}$: gravitational torque of the core on the shell.
\end{itemize}
\citet{bvyk2011} have shown that the sum of the torques acting on the ocean is null if the fluid is in hydrostatic equilibrium.

\par Since we work in the non-inertial reference frames of the principal axes of inertia of the core $(\vec{f_1^c},\vec{f_2^c},\vec{f_3^c})$, and of the shell
$(\vec{f_1^s},\vec{f_2^s},\vec{f_3^s})$, we must add $-\vec{\omega^c}\times\vec{G^c}$ in Eq.(\ref{eq:dGc}) and $-\vec{\omega^s}\times\vec{G^s}$ in 
Eq.(\ref{eq:dGs}).

We now detail the torques affecting the rotation.

\subsection{The gravitational pull of Saturn}

\par For a rigid triaxial body whose principal moments of inertia are $A<B<C$, the gravitational torque of a perturber is classically given by:

\begin{equation}
	\label{eq:pullrigid}
	\vv{\Gamma} = \frac{3\mathcal{G}M_{\saturn}}{\|\vv{r_{\saturn}}\|^5}\left((C-B)y_{\saturn}z_{\saturn}\vec{f_1}+(A-C)x_{\saturn}z_{\saturn}\vec{f_2}+(B-A)x_{\saturn}y_{\saturn}\vec{f_3}\right),
\end{equation}
where $\mathcal{G}$ is the gravitational constant, $M_{\saturn}$ the mass of the perturber, and the vector 
$\vv{r_{\saturn}}=x_{\saturn}\vec{f_1}+y_{\saturn}\vec{f_2}+z_{\saturn}\vec{f_3}$ locates the perturber from the center of mass of the triaxial body. 
In our case, this is the Titan-Saturn vector. A derivation of this torque is proposed in \citep{md2000}, inspired from \citep{m1936,r1937,r1940}.

\par We here treat the inner core and the shell as 2 independent triaxial rigid bodies, and we have:

\begin{eqnarray}
	\vv{\Gamma^c_{\saturn}} & = & \frac{3\mathcal{G}M_{\saturn}}{\|\vv{r^c_{\saturn}}\|^5}\left((C^c-B^c)y^c_{\saturn}z^c_{\saturn}\vv{f^c_1}+(A^c-C^c)x^c_{\saturn}z^c_{\saturn}\vv{f^c_2}+(B^c-A^c)x^c_{\saturn}y^c_{\saturn}\vv{f^c_3}\right), \label{eq:pullcore} \\
	\vv{\Gamma^s_{\saturn}} & = & \frac{3\mathcal{G}M_{\saturn}}{\|\vv{r^s_{\saturn}}\|^5}\left((C^s-B^s)y^s_{\saturn}z^s_{\saturn}\vv{f^s_1}+(A^s-C^s)x^s_{\saturn}z^s_{\saturn}\vv{f^s_2}+(B^s-A^s)x^s_{\saturn}y^s_{\saturn}\vv{f^s_3}\right). \label{eq:pullshell}
\end{eqnarray}

\par We use the TASS1.6 ephemerides \citep{vd1995} to express the Titan-Saturn vector $\vec{r_{\saturn}}$. These ephemerides are presented under a quasiperiodic form,
in which every sinusoidal contribution due to any perturber like Saturn's oblateness or the Solar attraction is explicitly expressed. They are given in the
inertial reference frame defined by the equatorial plane of Saturn at J2000 and the node of this plane with the ecliptic at the same date. As an example,
in this reference frame, the quantities related to the inclination of Titan $I_6$ and its ascending $\ascnode_6$ (Titan is denoted as \emph{S-6 Titan}) are

\begin{equation}
	\label{eq:inclitass}
	\sin\left(\frac{I_6(t)}{2}\right)\exp\left(\imath\ascnode_6(t)\right) = \gamma_0\exp\left(\imath\Omega_0\right)+\sum_{i=1}^4\gamma_i\exp\left(\imath\Omega_i(t)\right),
\end{equation}
the quantities $\gamma_i$ and $\Omega_i$ being given in the Tab.\ref{tab:inclitass}.

\begin{table}[ht]
	\centering
	\caption[Orbital inclination and ascending node of Titan.]{Relevant quantities for the orbital inclination and ascending node of Titan, reproduced from \citep{vd1995}. These numbers are to be used
	into the Eq.\ref{eq:inclitass}. The time origin is here J2000.\label{tab:inclitass}}
	\begin{tabular}{lrrrr}
		\hline
		i & Amplitude & Phase   & $d\Omega_i/dt$ & Period \\
		&   $\gamma_i$   &  $\Omega_i(t=0)$ & (rad/y)        & (years) \\
		\hline
		0 & $5.6024\times 10^{-3}$ & $184.578^{\circ}$ & 0 & -- \\
		1 & $2.7899\times 10^{-3}$ & $-14.731^{\circ}$ & $-8.93124\times10^{-3}$ &  $703.51$ \\
		2 & $1.312\times 10^{-4}$  & $-73.192^{\circ}$ & $-1.92554\times10^{-3}$ & $3263.07$ \\
		3 & $1.126\times 10^{-4}$  & $117.445^{\circ}$ & $0.42659824$ & $14.73$ \\
		4 & $1.92\times 10^{-5}$   &  $47.498^{\circ}$ & $-0.21329912$ & $29.46$ \\
		\hline
	\end{tabular}
\end{table}
Similar tables exist for the mean motion, the mean longitude, and the quantity relevant to the eccentricity $e_6$ and the longitude of 
the pericenter $\varpi_6$, i.e. $z_6=e_6\exp\left(\imath\varpi_6\right)$.

\par The TASS1.6 ephemerides are probably not the most accurate we could get, but they have the huge advantage of being presented in a quasiperiodic 
form giving explicitly the different contributions affecting Titan's orbit. From the periods of the sinusoidal quantities indexed from 1 to 4 in the 
Eq.(\ref{eq:inclitass}) and the Tab.\ref{tab:inclitass}, we can say that $i=1$ corresponds to the motion of the ascending node induced by the oblateness of 
Saturn, $i=2$ is a perturbation by Iapetus, and $i=3,4$ correspond to the Solar orbital perturbation, the orbital period of Saturn around the Sun being 
$29.46$ years. It is impossible to have such a decomposition with the JPL HORIZON ephemerides since they are given over a too short timespan 
($\approx400$ years) with respect to the relevant periods. This is why we choose to use TASS1.6, \citet{bvyk2011} having made the same choice.

\par The choice of the reference frame is not straightforward. To get an obliquity having a straightforward physical meaning, it is often advisable to use the 
Laplace Plane, that minimizes the variations of the inclination. Unfortunately, there are in the literature several inconsistent definitions of the Laplace
Plane, since there are several ways to minimize the variations of the inclination 
(over which time interval should we minimize? how do we measure the variations of the inclination?\ldots). \citet{n2009} suggests using the constant term 
in the quasiperiodic decomposition of the inclination to define the inertial reference frame. This is a kind of averaging of the orbital plane, that is very close to 
the Laplace Plane. In particular, this choice avoids a problem of apparent erratic behavior of the rotation pole that could happen for a satellite orbiting
far off its parent planet, when the rings' plane is chosen as the reference plane. The reason is that these satellites usually have a significantly inclined 
orbit because of the Solar perturbation.

\par This is why our reference frame is obtained from the reference frame of the ephemerides after 2 rotations: a $R_3$ rotation of $184.578^{\circ}$, and a 
$R_1$ rotation of $0.642^{\circ}$, these numbers being derived from the line $i=0$ in the Tab.\ref{tab:inclitass}. A very easy way to implement these 
rotations is just to drop $\gamma_0\exp\left(\imath\Omega_0\right)$ from Eq.(\ref{eq:inclitass}).

\par After a straightforward calculation we get from TASS1.6 the Titan-Saturn vector $\vec{r_0}$ in the inertial reference frame that we just defined, and
then the vector $\vv{r^{c,s}_{\saturn}}$ after 3 rotations of the Euler angles:

\begin{eqnarray}
	\vv{r^c_{\saturn}} & = & R_3(-\theta^c)R_1(-\epsilon^c)R_3(-h^c)\vec{r_0}, \label{eq:rotrc} \\
	\vv{r^s_{\saturn}} & = & R_3(-\theta^s)R_1(-\epsilon^s)R_3(-h^s)\vec{r_0}.
\end{eqnarray}

\par At this time we can simulate the rotation of the inner core and the rigid shell as two independent rigid bodies. We successfully checked our code in 
comparing with the results given by the numerical code based on an Hamiltonian formalism that we used in several previous studies \citep{nlv2008,n2009,n2010,nkr2011}.

\subsection{The gravitational coupling between the inner core and the shell\label{sec:gravicoup}}

\par This subsection is based on \citet{sx1997}, in which the gravitational coupling between the shell and the core of the Earth is estimated. 
This gravitational coupling is due to the misalignment of the principal axes of inertia of these 2 layers. We now need to be more specific as to their structure:

\begin{itemize}
	
\item The inner core is a triaxial ellipsoid with a constant density $\rho_c$, its radii being denoted $c_c < b_c < a_c$. Its surface is the core-ocean 
	boundary.
	
\item The shell (or crust) has a constant density $\rho_s$ as well. Its shape can be described by two concentric and coaxial triaxial ellipsoids. The radii of 
	the outer one are denoted $c < b < a$, they correspond to the observed shape of Titan \citep{zshlkl2009}, while the inner one is aligned with the outer one, its radii are 
	denoted $c_o < b_o < a_o$. This inner edge of the crust is the shell-ocean boundary. It is important to define $a_o$,$b_o$,$c_o$ as well as $a$,$b$ and $c$ to 
	allow for the possibility of lateral shell thickness variations,  which \citet{nb2010} argued are likely to exist.
	
\end{itemize}

\par Following \citet{sx1997}, the gravitational torque acting on the inner core due to the shell reads:

\begin{equation}
	\label{eq:torquesx}
	\vv{\Gamma^c_{sh}} = \iiint_{core}\rho_c \vec{r}\times\vec{\nabla}\Phi dV,
\end{equation}
where $\vec{r}$ points in this subsection to the position of a mass element of the core, and $\Phi$ is the potential of the shell given by \citep{sx1997}:

\begin{equation}
	\label{eq:phisx}
	\Phi = \alpha+\beta r^2P_2\left(\cos\psi\right)+\gamma r^2P_2^2\left(\cos\psi\right)\cos 2\phi+\mathcal{O}(f_{1,2}^2,\kappa_{1,2}^2),
\end{equation}
where

\begin{itemize}

	\item $\alpha$ is a constant that does not affect the result,
	
	\item $\beta=\frac{4\pi}{15}\mathcal{G}\rho_s\left(\kappa_2-\kappa_1-2(f_2-f_1)\right)$,
		
	\item $\gamma=\frac{2\pi}{15}\mathcal{G}\rho_s(\kappa_2-\kappa_1)$,
		
	\item $f_1=(a_o-c_o)/a_o$ is the polar flattening of the shell-ocean boundary,
		
	\item $f_2=(a-c)/a$ is the polar flattening of the surface of Titan,
		
	\item $\kappa_1=(a_o-b_o)/a_o$ is the equatorial ellipticy of the shell-ocean boundary,
		
	\item $\kappa_2=(a-b)/a$ is the equatorial ellipticity of the surface of Titan,
		
	\item $P_2(x)=(3x^2-1)/2$ is a Legendre polynomial,
		
	\item $P_2^2(x)=3(1-x^2)$ is a Legendre associated function,
		
	\item $\psi$ and $\phi$ are respectively the colatitude and the east longitude of the mass element involved, in the reference frame of the principal
		axes of inertia of the shell $\left(\vv{f_1^s},\vv{f_2^s},\vv{f_3^s}\right)$.
		
\end{itemize}


\par After some algebra (see App.\ref{sec:szetoxu}) we get

\begin{eqnarray}
	\label{eq:torquesx3}
	\vv{\Gamma^c_{sh}} & = & 3\beta \left[\begin{array}{c}
	n_2n_3(C^c-B^c) \vv{f_1^c} \\
	-n_1n_3(C^c-A^c) \vv{f_2^c} \\
	n_1n_2(B^c-A^c) \vv{f_3^c} \end{array}\right] + 6\gamma\left[\begin{array}{c}
	(l_2l_3-m_2m_3)(C^c-B^c) \vv{f_1^c} \\
	-(l_1l_3-m_1m_3)(C^c-A^c) \vv{f_2^c} \\
	(l_1l_2-m_1m_2)(B^c-A^c) \vv{f_3^c} \end{array}\right],
\end{eqnarray}	
where $l_i$, $m_i$ and $n_i$ ($i=1,2,3$) are the elements of the transition matrix between the coordinates in the reference frame of the shell $(X,Y,Z)$ and the ones
in the reference frame of the core $(x,y,z)$, i.e.

\begin{equation}
	\label{eq:lmn3}
	\left(\begin{array}{c}
	X \\
	Y \\
	Z \end{array}\right) = \left(\begin{array}{ccc}
	l_1 & l_2 & l_3 \\
	m_1 & m_2 & m_3 \\
	n_1 & n_2 & n_3 \end{array}\right)\left(\begin{array}{c}
	x \\
	y \\
	z \end{array}\right).
\end{equation}

\par An analogous calculation gives us the torque of the inner core acting on the shell:

\begin{equation}
	\label{eq:torqueshell}
	\vv{\Gamma^s_{co}}  =  -3\beta \left[\begin{array}{c}
	n_2n_3(C^c-B^c) \vv{f_1^s} \\
	-n_1n_3(C^s-A^c) \vv{f_2^s} \\
	n_1n_2(B^c-A^c) \vv{f_3^s} \end{array}\right] - 6\gamma\left[\begin{array}{c}
	(l_2l_3-m_2m_3)(C^c-B^c) \vv{f_1^s} \\
	-(l_1l_3-m_1m_3)(C^c-A^c) \vv{f_2^s} \\
	(l_1l_2-m_1m_2)(B^c-A^c) \vv{f_3^s} \end{array}\right].
\end{equation}

\subsection{Influence of the ocean}

\par \citet{bvyk2011} have shown that if the ocean is in hydrostatic equilibrium, then the pressure torque can be expressed as additional terms in the 
gravitational torques acting on the two rigid layers. We can split the ocean into two parts: a top and a bottom one, their boundary being spherical. The 
resulting pressure torque integrated over a spherical boundary is null, so the radius of this sphere has no influence on the results (an outcome we verified 
numerically). 

\par We call $A^o_t$, $B^o_t$, $C^o_t$, $A^o_b$, $B^o_b$ and $C^o_b$ the principal moments of inertia, respectively of the top ocean in the reference frame
of the shell $(\vv{f_1^s},\vv{f_2^s},\vv{f_3^s})$ and of the bottom ocean in the reference frame of the core $(\vv{f_1^c},\vv{f_2^c},\vv{f_3^c})$. Instead of 
writing the contribution of the ocean as independent torques $\vv{\Gamma^c_o}$ and $\vv{\Gamma^s_o}$, it is more appropriate to alter the other torques as
$\vv{\Gamma^{c,o}_{\saturn}}$, $\vv{\Gamma^{s,o}_{\saturn}}$, $\vv{\Gamma^{c,o}_{sh}}$, and $\vv{\Gamma^{s,o}_{co}}$. And we have:

\begin{equation}
	\label{eq:pulloccore}
	\begin{split}
	\vv{\Gamma^{c,o}_{\saturn}}  = \frac{3\mathcal{G}M_{\saturn}}{\|\vv{r^c_{\saturn}}\|^5}((C^c-B^c+C^o_b-B^o_b)y^c_{\saturn}z^c_{\saturn}\vv{f^c_1}+(A^c-C^c+A^o_b-C^o_b)x^c_{\saturn}z^c_{\saturn}\vv{f^c_2} \\
	 +  (B^c-A^c+B^o_b-A^o_b)x^c_{\saturn}y^c_{\saturn}\vv{f^c_3}),
	\end{split}
\end{equation}

\begin{equation}
	\label{eq:pullocshell}
	\begin{split}
	\vv{\Gamma^{s,o}_{\saturn}}  =  \frac{3\mathcal{G}M_{\saturn}}{\|\vv{r^s_{\saturn}}\|^5}((C^s-B^s+C^o_t-B^o_t)y^s_{\saturn}z^s_{\saturn}\vv{f^s_1}+(A^s-C^s+A^o_t-C^o_t)x^s_{\saturn}z^s_{\saturn}\vv{f^s_2} \\
	 +  (B^s-A^s+B^o_t-A^o_t)x^s_{\saturn}y^s_{\saturn}\vv{f^s_3}), 
	\end{split}
\end{equation}

%

\begin{equation}
	\label{eq:torqueocinner}
	\vv{\Gamma^{c,o}_{sh}}  =  \left[\begin{array}{c}
	\left(3\beta^o n_2n_3+6\gamma^o(l_2l_3-m_2m_3)\right)(C^c-B^c+C^o_b-B^o_b) \vv{f_1^c} \\
	-\left(3\beta^on_1n_3+6\gamma^o(l_1l_3-m_1m_3)\right) (C^c-A^c+C^o_b-A^o_b) \vv{f_2^c} \\
	\left(3\beta^on_1n_2+6\gamma^o(l_1l_2-m_1m_2)\right)(B^c-A^c+B^o_b-A^o_b) \vv{f_3^c} \end{array}\right],
\end{equation}	

and

\begin{equation}
	\label{eq:torqueocshell}
	\vv{\Gamma^{s,o}_{co}}  =  \left[\begin{array}{c}
	-\left(3\beta^o n_2n_3+6\gamma^o(l_2l_3-m_2m_3)\right)(C^c-B^c+C^o_b-B^o_b) \vv{f_1^s} \\
	\left(3\beta^on_1n_3+6\gamma^o(l_1l_3-m_1m_3)\right) (C^c-A^c+C^o_b-A^o_b) \vv{f_2^s} \\
	-\left(3\beta^on_1n_2+6\gamma^o(l_1l_2-m_1m_2)\right)(B^c-A^c+B^o_b-A^o_b) \vv{f_3^s} \end{array}\right],
\end{equation}	
with

\begin{eqnarray}
	\beta^o  & = & \frac{4\pi}{15}\mathcal{G}\left(\rho_s\left(\kappa_2-\kappa_1-2(f_2-f_1)\right)+\rho_o\left(\kappa_1-2f_1\right)\right), \label{eq:betao} \\
	\gamma^o & = & \frac{2\pi}{15}\mathcal{G}\left(\rho_s(\kappa_2-\kappa_1)+\rho_o\kappa_1\right), \label{eq:gammao}
\end{eqnarray}
$\rho_o$ being the constant density of the ocean.

\par We have now the whole equations of the problem, consisting of 12 variables $\xi^c$, $\eta^c$, $p^c$, $\omega_1^c$, $\omega_2^c$, $\omega_3^c$, $\xi^s$, 
$\eta^s$, $p^s$, $\omega_1^s$, $\omega_2^s$, $\omega_3^s$, the first 6 describing the orientation of the core, and the last 6 the orientation of the shell.
The components of the rotation vector $\vec{\omega}$ in the reference frame of the principal axes of inertia of the considered layer are obtained from the 
angular momentun $\vec{G}$ and division by the appropriate moment of inertia.

\par As a summary, we here gather these equations. We have:

\begin{eqnarray}
  	\frac{d\xi^c}{dt}      & = & \left(\omega_1^c\cos\theta^c-\omega_2^c\sin\theta^c\right)\sin h^c+\frac{\epsilon^c}{\sin\epsilon^c}\left(\omega_1^c\sin\theta^c+\omega_2^c\cos\theta^c\right)\cos h^c, \label{eq:dotxic} \\
	\frac{d\eta^c}{dt}     & = & \left(\omega_1^c\cos\theta^c-\omega_2^c\sin\theta^c\right)\cos h^c-\frac{\epsilon^c}{\sin\epsilon^c}\left(\omega_1^c\sin\theta^c+\omega_2^c\cos\theta^c\right)\sin h^c, \label{eq:dotetac} \\
	\frac{dp^c}{dt}        & = & \omega_3^c+\left(\omega_1^c\sin\theta^c+\omega_2^c\cos\theta^c\right)\tan\frac{\epsilon^c}{2}, \label{eq:dotpc} \\
	\frac{d\omega_1^c}{dt} & = & \left(3\beta^on_2n_3+6\left(l_2l_3-m_2m_3\right)\right)\frac{C^c-B^c+C_b^o-B_b^o}{A^c} \nonumber \\
	                       & + & 3\frac{\mathcal{G}M_{\saturn}}{||\vv{r_{\saturn}^c}||^5}\frac{C^c-B^c+C_b^o-B_b^o}{A^c}y_{\saturn}^cz_{\saturn}^c-\frac{\left(\vec{\omega^c}\times\vec{G^c}\right)\cdot\vv{f_1^c}}{A^c}, \label{eq:dotomega1c} \\
	\frac{d\omega_2^c}{dt} & = & -\left(3\beta^on_1n_3+6\left(l_1l_3-m_1m_3\right)\right)\frac{C^c-A^c+C_b^o-A_b^o}{B^c} \nonumber \\
	                       & + & 3\frac{\mathcal{G}M_{\saturn}}{||\vv{r_{\saturn}^c}||^5}\frac{A^c-C^c+A_b^o-C_b^o}{B^c}x_{\saturn}^cz_{\saturn}^c-\frac{\left(\vec{\omega^c}\times\vec{G^c}\right)\cdot\vv{f_2^c}}{B^c}, \label{eq:dotomega2c} \\
	\frac{d\omega_3^c}{dt} & = & \left(3\beta^on_1n_2+6\left(l_1l_2-m_1m_2\right)\right)\frac{B^c-A^c+B_b^o-A_b^o}{C^c} \nonumber \\
	                       & + & 3\frac{\mathcal{G}M_{\saturn}}{||\vv{r_{\saturn}^c}||^5}\frac{B^c-A^c+B_b^o-A_b^o}{C^c}x_{\saturn}^cy_{\saturn}^c-\frac{\left(\vec{\omega^c}\times\vec{G^c}\right)\cdot\vv{f_3^c}}{C^c} \label{eq:dotomega3c}
\end{eqnarray}
for the core, and 

\begin{eqnarray}
	\frac{d\xi^s}{dt}      & = & \left(\omega_1^s\cos\theta^s-\omega_2^s\sin\theta^s\right)\sin h^s+\frac{\epsilon^s}{\sin\epsilon^s}\left(\omega_1^s\sin\theta^s+\omega_2^s\cos\theta^s\right)\cos h^s, \label{eq:dotxis} \\
	\frac{d\eta^s}{dt}     & = & \left(\omega_1^s\cos\theta^s-\omega_2^s\sin\theta^s\right)\cos h^s-\frac{\epsilon^s}{\sin\epsilon^s}\left(\omega_1^s\sin\theta^s+\omega_2^s\cos\theta^s\right)\sin h^s, \label{eq:dotetas} \\
	\frac{dp^s}{dt}        & = & \omega_3^s+\left(\omega_1^s\sin\theta^s+\omega_2^s\cos\theta^s\right)\tan\frac{\epsilon^s}{2}, \label{eq:dotps} \\
	\frac{d\omega_1^s}{dt} & = & -\left(3\beta^on_2n_3+6\left(l_2l_3-m_2m_3\right)\right)\frac{C^c-B^c+C_b^o-B_b^o}{A^s} \nonumber \\
	                       & + & 3\frac{\mathcal{G}M_{\saturn}}{||\vv{r_{\saturn}^s}||^5}\frac{C^s-B^s+C_t^o-B_t^o}{A^s}y_{\saturn}^sz_{\saturn}^s-\frac{\left(\vec{\omega^s}\times\vec{G^s}\right)\cdot\vv{f_1^s}}{A^s}, \label{eq:dotomega1s} \\
	\frac{d\omega_2^s}{dt} & = & \left(3\beta^on_1n_3+6\left(l_1l_3-m_1m_3\right)\right)\frac{C^c-A^c+C_b^o-A_b^o}{B^s} \nonumber \\
	                       & + & 3\frac{\mathcal{G}M_{\saturn}}{||\vv{r_{\saturn}^s}||^5}\frac{A^s-C^s+A_t^o-C_t^o}{B^s}x_{\saturn}^sz_{\saturn}^s-\frac{\left(\vec{\omega^s}\times\vec{G^s}\right)\cdot\vv{f_2^s}}{B^s}, \label{eq:dotomega2s} \\
	\frac{d\omega_3^s}{dt} & = & -\left(3\beta^on_1n_2+6\left(l_1l_2-m_1m_2\right)\right)\frac{B^c-A^c+B_b^o-A_b^o}{C^s} \nonumber \\
	                       & + & 3\frac{\mathcal{G}M_{\saturn}}{||\vv{r_{\saturn}^s}||^5}\frac{B^s-A^s+B_t^o-A_t^o}{C^s}x_{\saturn}^sy_{\saturn}^s-\frac{\left(\vec{\omega^s}\times\vec{G^s}\right)\cdot\vv{f_3^s}}{C^s} \label{eq:dotomega3s}
\end{eqnarray}
for the shell.

\clearpage

\section{A numerical solution}

\par A numerical solution of the equations (\ref{eq:dotxic}) to (\ref{eq:dotomega3s}) is here appropriate since we want to include complete ephemerides and 6 dynamical degrees of freedom.

\subsection{Numerical integration of the equations}

\par The numerical integrations are performed with the Adams-Bashforth-Moulton 10th order predictor corrector integrator (see e.g. \citep{hnw1993}), with a tolerance
of $10^{-14}$ and a step size of $0.2$ day. This corresponds to $\approx1/80$ of the orbital period of Titan.

\par The rotation of Titan is expected to be at a dynamical equilibrium. Such equilibriums are known as Cassini States \citep{c1693,c1966,p1969} for rigid bodies. The
expected state for the natural satellites of the giant planets is Cassini State 1 since it is the most stable. In our case of a 3-layer Titan, we initially assume an 
analogous state in which the inner core and the crust are close to the location of Cassini State 1 if they were only interacting with Saturn. This state corresponds 
to a synchronous rotation, a small obliquity and a small polar motion. As a consequence, the angular momentum of the shell and the core should approximate $\vv{G^s}\approx n_6C^s \vv{f_3^s}$ and
$\vv{G^c}\approx n_6C^c \vv{f_3^c}$ where $n_6$ is the mean motion, or orbital frequency, of Titan, and the spin angles of the core $p^c$ and of the shell $p^s$ should be always close to 
the orbital mean longitude of Titan $\lambda_6$.

\par Because of the effects that are neglected in the theory of the Cassini States, especially the couplings between the different layers and the perturbations considered in the 
orbital motion of Titan, it is very difficult to derive analytically the optimal initial conditions for the Euler angles and the rotation vector. In practice, our initial conditions are usually close enough to the optimum state that oscillations round the equilibrium result, these free oscillations having an arbitrary amplitude due to the choice of the initial conditions, and a proper frequency whose value depends on the
parameters of the system, here the interior of Titan. These free oscillations pollute the analysis of the solutions in acting as a noise, for this reason we wish their amplitude to be as 
small as possible. For that, we refine numerically the initial conditions thanks to an iterative algorithm based on the frequency analysis.


\par The basic idea is that since the orbital motion of Titan can be given under a quasiperiodic form, and that the rotation of Titan is not expected to be chaotic, then this rotation
can be expressed under a quasiperiodic form as well. A complex variable $x(t)$ of the problem that does not diverge can read as a sum of a converging trigonometric series like

\begin{equation}
 x(t)=\sum_{n=0}^{\infty} A_n \exp\left(\imath \nu_nt\right),
\end{equation}
where $A_n$ are constant complex amplitudes, and $\nu_n$ constant frequencies, with 

\begin{equation}
 x(t)\approx\sum_{n=0}^{N} A_n^{\bullet} \exp\left(\imath \nu_n^{\bullet}t\right),
\label{eq:naffc}
\end{equation}
the bullet meaning that the coefficients have been numerically determined. A detailed description of the algorithm is given in 
Appendix B.  In the case of a real variable, Eq.\ref{eq:naffc} becomes

\begin{equation}
 x(t)\approx\sum_{n=0}^{N} A_n^{\bullet} \cos\left(\nu_n^{\bullet}t+\phi_n^{\bullet}\right),
\label{eq:naffrc}
\end{equation}
or

\begin{equation}
 x(t)\approx\sum_{n=0}^{N} A_n^{\bullet} \sin\left(\nu_n^{\bullet}t+\phi_n^{\bullet}\right),
\label{eq:naffrs}
\end{equation}
where the amplitudes are now real, and the $\phi_n^{\bullet}$ are real phases, previously included in the complex amplitudes in Eq.\ref{eq:naffc}.

The frequency analysis algorithm we use is based on NAFF (see \citep{l1993} for the method, and \citep{l2005} for the convergence proofs), with a refinement suggested by 
\citep{c1998} consisting in iterating the process to improve the accuracy of the determination. The frequencies $\nu_n$ have 2 origins: they might be either forcing frequencies, present 
in the orbital motion of Titan, or free frequencies, due to the departure from the exact equilibrium. The amplitude associated with the latter should be as small as possible. To get the appropriate
initial conditions we use an iterative algorithm \citep{ndc2014}, consisting in:

\begin{enumerate}

 \item A first numerical integration of the equations of the system, with initial conditions conveniently chosen,

 \item Frequency analysis of the solution and identification of the contributions depending on the free modes,

 \item Evaluation of the free modes at the origin time of the numerical simulation, and removal from the initial conditions,

\end{enumerate}
then the process is iterated until convergence. This algorithm has already been successfully applied in problem of rotational dynamics \citep{dnrl2009,n2009,rrc2011}, 
in dynamics of exoplanetary systems \citep{clcmu2010}, and in the analysis of ground-track resonances around Vesta \citep{d2011}.

\par We know of at least 3 alternative methods to reduce the amplitude of the free librations:

\begin{itemize}

  \item \citet{br2007} propose to fit the mean initial conditions in order to locate the spin-orbit system at its center of libration,
  
  \item \citet{pym2007} add a damping in the equations that reduces the amplitude of the free librations. The damping must be slow enough, i.e. adiabatic,
  to not alter significantly the location of the equilibrium,
  
  \item \citet{ym2006}, in the framework of a numerical integration of the spin and of the orbit of Mercury, start from a simple Sun-Mercury system 
  in which the equilibrium is very easy to determine analytically, and slowly switch on the planetary perturbations in order to create an adiabatic
  devitation of the equilibrium without creation of any free libration. In our case, this would require us to simultaneously integrate the orbit of Titan, rather than use the existing ephemerides.

\end{itemize}
All of these methods, including ours, give accurate results when appropriately used.

\par In this study, we simulated the rotation of thousands of model Titans (see Sec.\ref{sec:interior}) and did not refine the initial conditions for all of them. In practice, 
we did it for just a few of them, and got initial conditions that we considered to be good enough for the remainder.

\subsection{Outputs}

\par Our set of variables describes all the dynamical degrees of freedom of the core and the shell, so we are able to express any observable of the rotation. Our outputs are, 
for these two rigid layers:

\begin{itemize}
 \item the longitudinal librations,
 \item the obliquity,
 \item and the polar motion.
\end{itemize}

\par The longitudinal librations of the shell are usually considered as the most significant output since they can reveal a global fluid layer \citep{vrkdr2008}. There are at least 
two ways to define them: the tidal librations and the physical librations. The tidal librations $\psi^{s,c}$ represent the longitudinal misalignment between the directions of the 
long axis of the layer under consideration (the shell or the core) and the Saturn-Titan direction. The physical significance of these librations is that they control the amount 
of tidal stress and heating arising. They are given by

\begin{eqnarray}
  \psi^s & = & \vec{t}\cdot\vec{f_1^s}, \label{eq:psis} \\
  \psi^c & = & \vec{t}\cdot\vec{f_1^c}, \label{eq:psic}
\end{eqnarray}
where $\vec{t}$ is the unit vector tangent to the trajectory of Titan around Saturn:

\begin{equation}
  \label{eq:tangent}
  \vec{t}=\frac{\vec{n}\times\vec{x}}{||\vec{n}\times\vec{x}||},
\end{equation}
and $\vec{n}$ the unit vector normal to the orbit:

\begin{equation}
  \label{eq:normale}
  \vec{n}=\frac{\vec{x}\times\vec{v}}{||\vec{x}\times\vec{v}||},
\end{equation}
$\vec{x}$ and $\vec{v}$ being respectively the position vector of Titan, and $\vec{v}$ its velocity.

The physical librations $\gamma^{s,c}$ are the librations about the exact synchronous rotation. We derived them from the variable $p^{s,c}-n_6t$, this is a very good approximation if the 
angles $\epsilon^{s,c}$ are small, i.e. if the two angles $h^s$ and $\theta^s$ constituting $p^s$, respectively $h^c$ and $\theta^c$ for $p^c$, are nearly coplanar. In practice, the 
angles $\epsilon^s$ and $\epsilon^c$ are always smaller than 1 degree. They also correspond, up to the first order in eccentricity, to the librations of the long axis of the  considered 
layer with the direction of the empty focus of the orbit of Titan.

The difference between the tidal and the physical librations are known as optical librations, their amplitude is twice the eccentricity $e_6$ and is just a signature of the orbital motion,
not of the interior of the body. In practice, the tidal librations are dominated by the optical librations. This is why the physical librations are usually preferred to the tidal ones, their
amplitude is roughly proportional to the difference of the moments of inertia (B-A) for rigid bodies.

In the case of Titan, longitudinal librations have not so far been detected, although analysis of {\em Cassini} radar images may ultimately make this possible \cite[e.g.]{bsk2013}.

The obliquity of the shell $K^s$ (respectively of the core $K^c$) is defined as the angle between the normale to the orbit $\vec{n}$ and the angular momentum of the shell $\vv{G^s}$ 
(respectively of the core $\vv{G^c}$). Since an obliquity belongs to the range $[0^{\circ}$-$180^{\circ}]$, we can define it with its cosine and we have:

\begin{eqnarray}
  K^s & = & \arccos\left(\frac{\vec{n}\cdot\vv{G^s}}{||\vv{G^s}||}\right), \label{eq:oblishell} \\
  K^c & = & \arccos\left(\frac{\vec{n}\cdot\vv{G^c}}{||\vv{G^c}||}\right). \label{eq:oblicore}
\end{eqnarray}

{\em Cassini} measured an obliquity of the shell of $0.31^{\circ}\pm0.05^{\circ}$, i.e. $(18.6\pm3)$ arcmin at the mean date March 11$^{th}$, 2007 \citep{mi2012}.

\par The polar motion is a priori expected to be very small and is often neglected. We did include polar motion in our numerical simulations for completeness, and in case 
resonances or nonlinearities in the rotational dynamics resulted in large polar motion, as suggested for a rigid Titan by \citep{n2008}, and for a Titan with a thin shell and a strong
atmospheric torque in \citep{tvk2011}. However, in every simulation polar motion remained very small.

\clearpage

\section{Possible interiors of Titan \label{sec:interior}}

\par The goal of this section is to build realistic models for Titan, that will give us the interior parameters we need in our numerical code, i.e. the density and the 3 outer
radii of the inner core, the ocean and the shell. The three external radii of the shell are known thanks to {\em Cassini} observations \citep{zshlkl2009}. To build our Titans,
our algorithm consists of 3 steps:

\begin{enumerate}

  \item Elaboration of hydrostatic Titans. The choice of this starting point comes from the observations that the gravity field of Titan is not far from a hydrostatic one,
  
  \item Modification of the hydrostatic state by including shell thickness variations, imposed at either the surface (top loading) or at the ocean-shell boundary (bottom loading),
  
  \item Comparison with the gravity field. Only model Titans for which the gravity field is consistent with the observations are retained.

\end{enumerate}

\par For each model Titan, we set the densities of the shell $\rho_s$ and of the ocean $\rho_o$, and the mean thicknesses of the shell $d_s$ and of the ocean $d_o$. The size and density of 
the core can then be deduced, the mean density of Titan being $\rho_6=1881$ $kg/m^3$ \citep{irjrstaa2010}. Note that in our approach the core can potentially include 
high-pressure phases of ice; the importance of the core is that it represents the base of the decoupling ocean. The range of internal parameters we consider are:

\begin{itemize}

  \item $d_s$ between $50$ and $200$ km,
  
  \item $d_o$ between $50$ and $400-d_s$ km,
  
  \item $\rho_s$ between $900$ and $950$ $kg/m^3$,
  
  \item $\rho_o$ between $950$ and $1200$ $kg/m^3$.

\end{itemize}

\par The range of possible densities for the ocean comes from \citet{f2012}. The limit of the depth of the core-ocean interface comes from the condition of existence
of a liquid ocean with respect to the temperature and the pressure, see the phase diagram of water ice \citep{smrss2010} and of an ammonia-water ocean \citep{shssl2003}.
Even for a warm ocean (270~K), the pressure at the base of the ocean  should be smaller than 0.6~GPa, which corresponds to a depth of about 450~km. Since ammonia is 
likely to be present, 270~K is likely an overestimate, so we consider that 400~km is a reasonable limit for the depth of the core-ocean boundary.

\par We here build initially hydrostatic Titans, following a method described in \citet{vrkdr2008}. A body in hydrostatic equilibrium has a shape corresponding to a 
balance between its own gravity, its rotation and the tidal deformation. Its surface is an equipotential,
and the 2 boundaries between the different layers are equipotential as well. We consider that the mean radius $R=2574.73$ km and the along-orbit equatorial radius $b=2574.78$ km 
are known. We then obtain from the Radau equation (see e.g. \citep{j1952}):

\begin{eqnarray}
h_2 & = & \frac{5}{1+\left(\frac{5}{2}-\frac{15}{4}\frac{C}{M_6R^2}\right)^2}, \label{eq:h2} \\
q & = & \frac{n_6^2R^3}{\mathcal{G}M_6}, \label{eq:q} \\ 
\tilde{c} & = & b-\frac{h_2}{2}qR, \label{eq:cradau} \\
\tilde{a} & = & b+\frac{3}{2}qh_2R, \label{eq:aradau}
\end{eqnarray}
where $h_2$ is the second-order fluid Love number related to the radial displacement. $\tilde{a}$ and $\tilde{c}$ are radii of this hydrostatic Titan, 
they should be very close to $a$ and $c$. $q=3.95428\times10^{-5}$ quantifies the relative influence between the 
rotation of Titan and its own gravity. $C=C^s+C^o+C^c$ is the polar moment of inertia of the whole Titan. In the Eq.(\ref{eq:h2}), $C/(M_6R^2)$ is in fact 
used as an approximation of $I/(M_6R^2)=(A+B+C)/(3M_6R^2)=C/(M_6R^2)-2J_2/3$. At this stage, the shape of Titan is unknown, so we neglect $J_2$
and we estimate $C$ from the mean radius R and the mean thicknesses of the shell $d_s$ and the ocean $d_o$, i.e.

\begin{equation}
  \label{eq:Capprox}
  C \approx \frac{8\pi}{15}\left(\left(\rho_c-\rho_o\right)\left(R-d_s-d_o\right)^5+\left(\rho_o-\rho_s\right)\left(R-d_s\right)^5+\rho_sR^5\right).
\end{equation}

\par The Radau equation comes from the solution of the Clairaut equation that gives the flattening $\alpha=(a+b-2c)/(a+b)$ of the equipotential surface at any radius $r$:

\begin{eqnarray}
  \frac{d^2\alpha}{dr^2}+\frac{6}{r}\frac{\rho}{\bar{\rho}}-\frac{6}{r^2}\left(1-\frac{\rho}{\bar{\rho}}\right)\alpha & = & 0, \label{eq:clairaut} \\
  \frac{d\alpha}{dr}(R) & = & \frac{1}{R}\left[\frac{25}{4}q-2\alpha(R)\right], \label{eq:clairautci1} \\
  \alpha(R) & = & \frac{\tilde{a}+b-2\tilde{c}}{\tilde{a}+b}. \label{eq:clairautci0}
\end{eqnarray}
The initial condition (\ref{eq:clairautci1}) is not the classical one given in the Clairaut theory where $25/4$ should be replaced by $5/2$, because it considers 
the deformations due to the rotation, and the tides. Moreover, we have $\beta=6\alpha/5=(\tilde{a}-b)/\tilde{a}$ at any radial distance $r$. This can be easily 
seen at $r=R$. From the classical relation $(\tilde{a}-\tilde{c})=4(b-\tilde{c})$ for hydrostatic synchronous bodies, we have straightforwardly 
$\alpha = 5/3\times(\tilde{a}-b)/(\tilde{a}+b)\approx5\beta/6$. A more rigorous proof, valid at any radial distance $r$, can be found in \citep{vrkdr2008}.
We then get the three radii of the core-ocean and of the ocean-shell boundaries.

\par These hydrostatic Titans do not correspond to the real one. In particular, the two model external radii $\tilde{a}$ and $\tilde{c}$ are not 
consistent with the observations. We solve this problem by introducing a topographic anomaly at the surface $h_t$:

\begin{equation}
  \label{eq:anomaly}
  h_t(\psi,\phi) = h_{t1}Y_{20}(\psi)+h_{t2}Y_{22}(\psi,\phi)
\end{equation}
where $Y_{20}$ and $Y_{22}$ are the classical second-degree spherical harmonics defined by:

\begin{equation}
  \label{eq:harmosferik}
  Y_{lm}(\psi,\phi) = \sqrt{\left(2-\delta_{0m}\right)(2l+1)\frac{(l-m)!}{(l+m)!}}P_{lm}(\cos\psi))\cos m\phi
\end{equation}
for $m>0$. $P_{l0}$ are the Legendre polynomials and $P_{lm}$ the associated Legendre functions for $m\ne0$ already defined in Subs.\ref{sec:gravicoup}. Eq.(\ref{eq:anomaly}) becomes

\begin{equation}
  \label{eq:anomaly2}
  h_t(\psi,\phi) = \frac{\sqrt{5}}{2}h_{t1}\left(3\cos^2\psi-1\right)+3\sqrt{\frac{5}{12}}h_{t2}\sin^2\psi\cos2\phi.
\end{equation}

We now set $\Delta a=a-\tilde{a}$  and $\Delta c=c-\tilde{c}$ the two radial anomalies corresponding respectively to $(\psi=k\pi)$ and $(\psi=k\pi,\phi=2k'\pi)$,
where $k$ and $k'$ are integers.
We have:

\begin{eqnarray}
  \Delta c & = & \sqrt{5}h_{t1}, \label{eq:deltac} \\
  \Delta a & = & -\frac{\sqrt{5}}{2}h_{t1}+3\sqrt{\frac{5}{12}}h_{t2}, \label{eq:deltaa}
\end{eqnarray}
from which we deduce $h_{t1}$ and $h_{t2}$.

It follows from the definition of the spherical harmonics that the radius $b$ is altered as well. For this reason, we need to iterate the process in correcting the initial value of $b$, 
so that the surface of our model Titan has the correct radii $a$, $b$ and $c$ at the end.

\par As suggested by \citet{nb2010,hnzi2013}, the topographic anomaly could be caused by variations of the thickness of the ice shell and partial or complete isostatic compensation (see e.g.\citep{w2001}). So, there should be a corresponding bottom anomaly $h_b$ altering the shape of the ocean-shell boundary following the same spherical harmonics, i.e.

\begin{equation}
  \label{eq:anomalyb}
  h_b(\psi,\phi) = h_{b1}Y_{20}(\psi)+h_{b2}Y_{22}(\psi,\phi).
\end{equation}

\par It is useful to distinguish between a load applied at the surface of Titan (top loading) and a load applied at the ocean-shell boundary (bottom loading). Top 
loading results in bottom topography $h_b$ given by:

\begin{equation}
  \label{eq:toploading}
  h_b = fh_t\left(\frac{R}{R-d_s}\right)^2\frac{\rho_s}{\rho_o-\rho_s},
\end{equation}
while bottom loading gives the following surface topography:

\begin{equation}
  \label{eq:bottomloading}
  h_t = fh_b\left(\frac{R-d_s}{R}\right)^2\frac{\rho_o-\rho_s}{\rho_s}.
\end{equation}

Here $f\in[0;1]$ is the compensation factor. $f=0$ means that there is no compensation, the shell is perfectly rigid, and loading does not cause any deformation of the shell. Contrariwise, 
$f=1$ means full compensation, so that the top- and bottom-loading cases are indistinguishable.

\par Once these model Titans have been computed, their second-degree gravity fields are calculated and only the ones consistent with the gravity solutions SOL1 or SOL2 are kept. 
Because the degree-3 gravity, not modelled here but treated in detail by \citet{hnzi2013}, is more consistent with bottom loading than with top loading, we focus more on the 
former. The only acceptable solutions we get with top loading lie between the $2\sigma$ and $3\sigma$ limits for SOL2. They are not displayed here since they cannot explain 
the observations (see Sect.\ref{sec:results}).

\par
\begin{figure}[ht]
\centering
\begin{tabular}{cc}
  \includegraphics[width=0.45\textwidth]{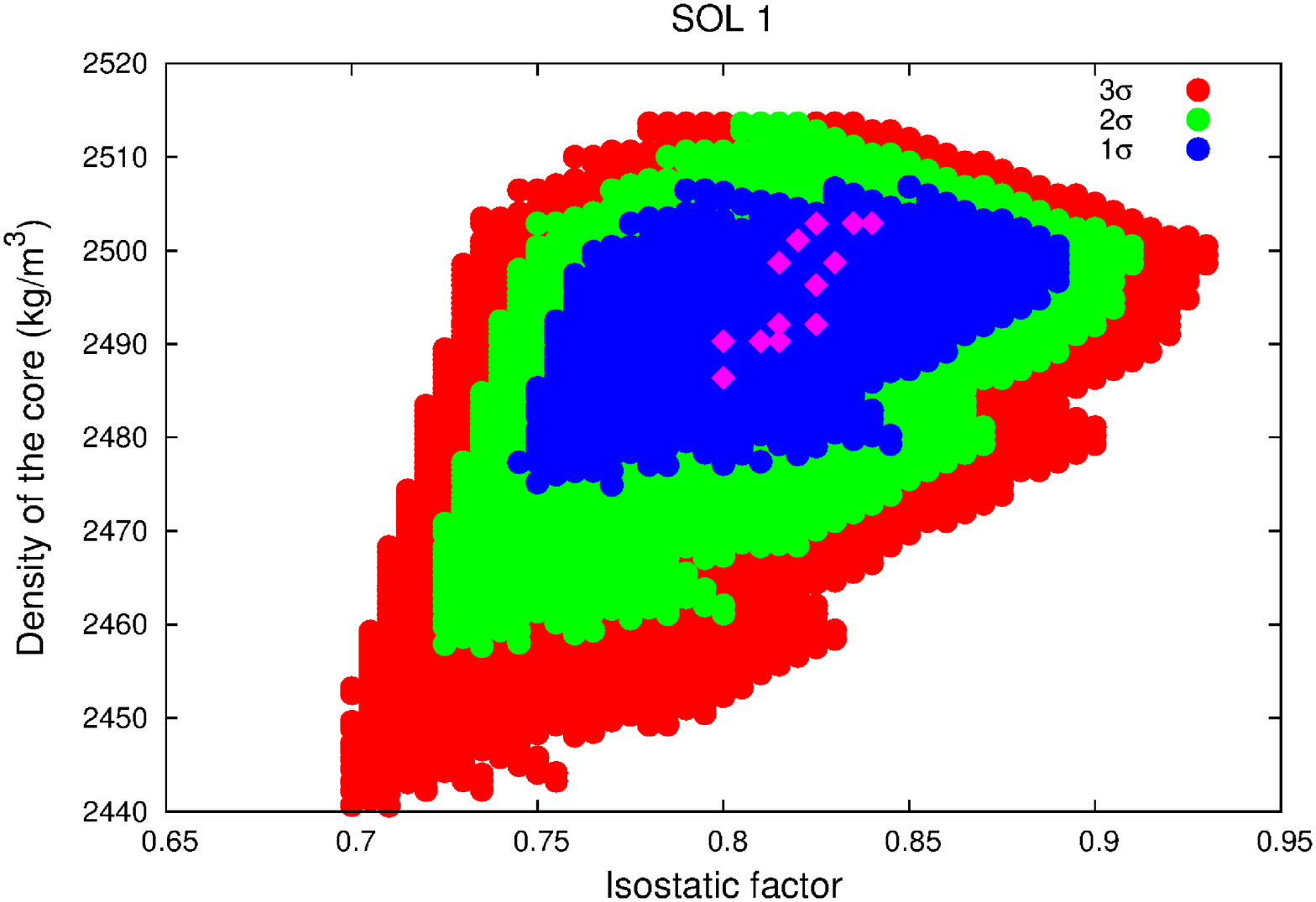} & \includegraphics[width=0.45\textwidth]{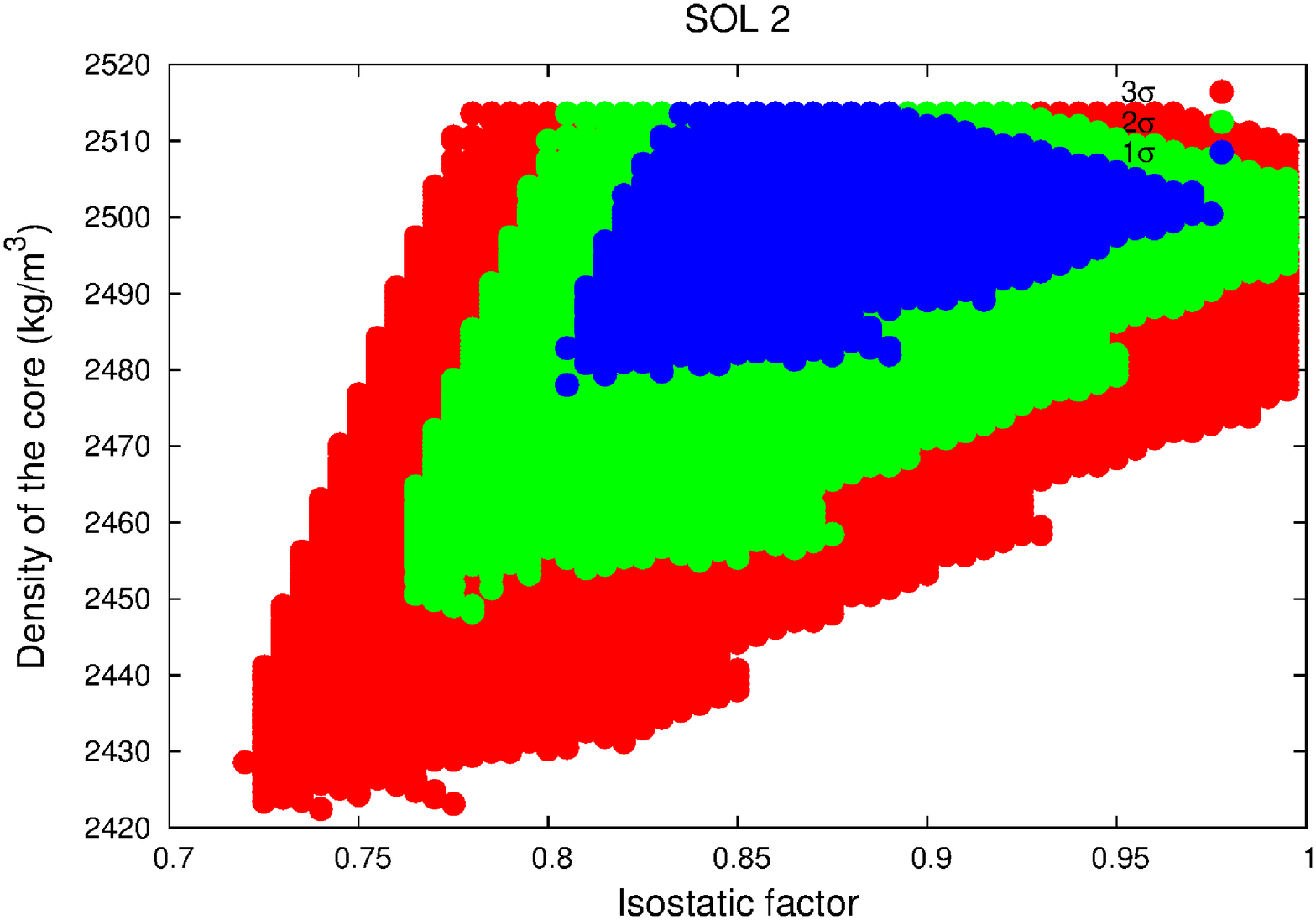}
\end{tabular}
\caption[Properties of our Titans.]{Properties of our model Titans, complying with the 2 possible gravity fields SOL1 (multi-arc) and SOL2 (global), at $1\sigma$ (blue), $2\sigma$ (green) and $3\sigma$ (red).
These 2 plots have been obtained assuming bottom loading. We can see in particular that SOL1 suggests a compensation factor between 75 and 88$\%$, while SOL2 suggests 
between 80 and 96$\%$ at $1\sigma$. The left panel also shows 13 model Titans that will be used later to simulate the behavior of the obliquity.\label{fig:iso2sol}}
\end{figure}

\begin{figure}[ht]
\centering
\begin{tabular}{cc}
  \includegraphics[width=0.45\textwidth]{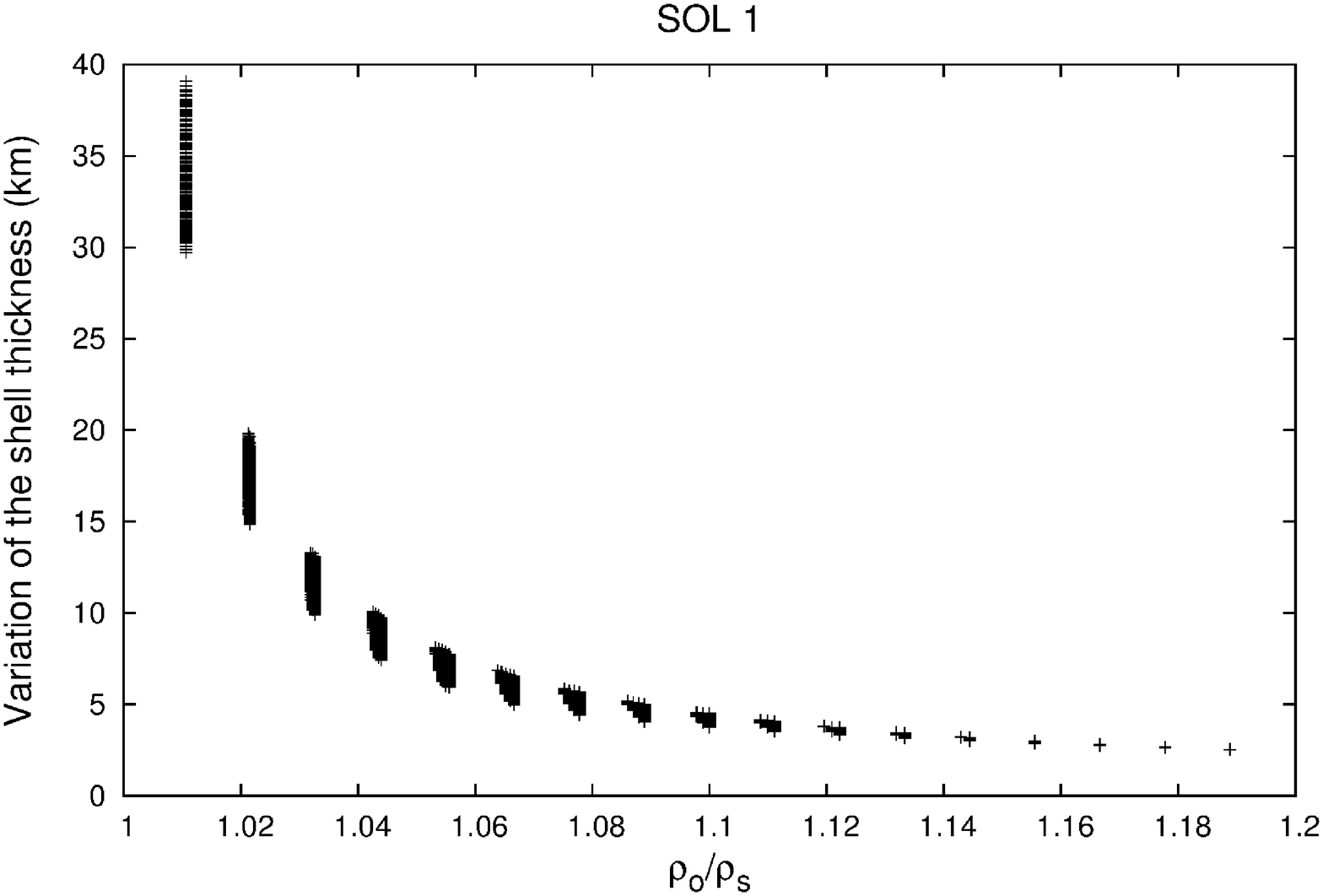} & \includegraphics[width=0.45\textwidth]{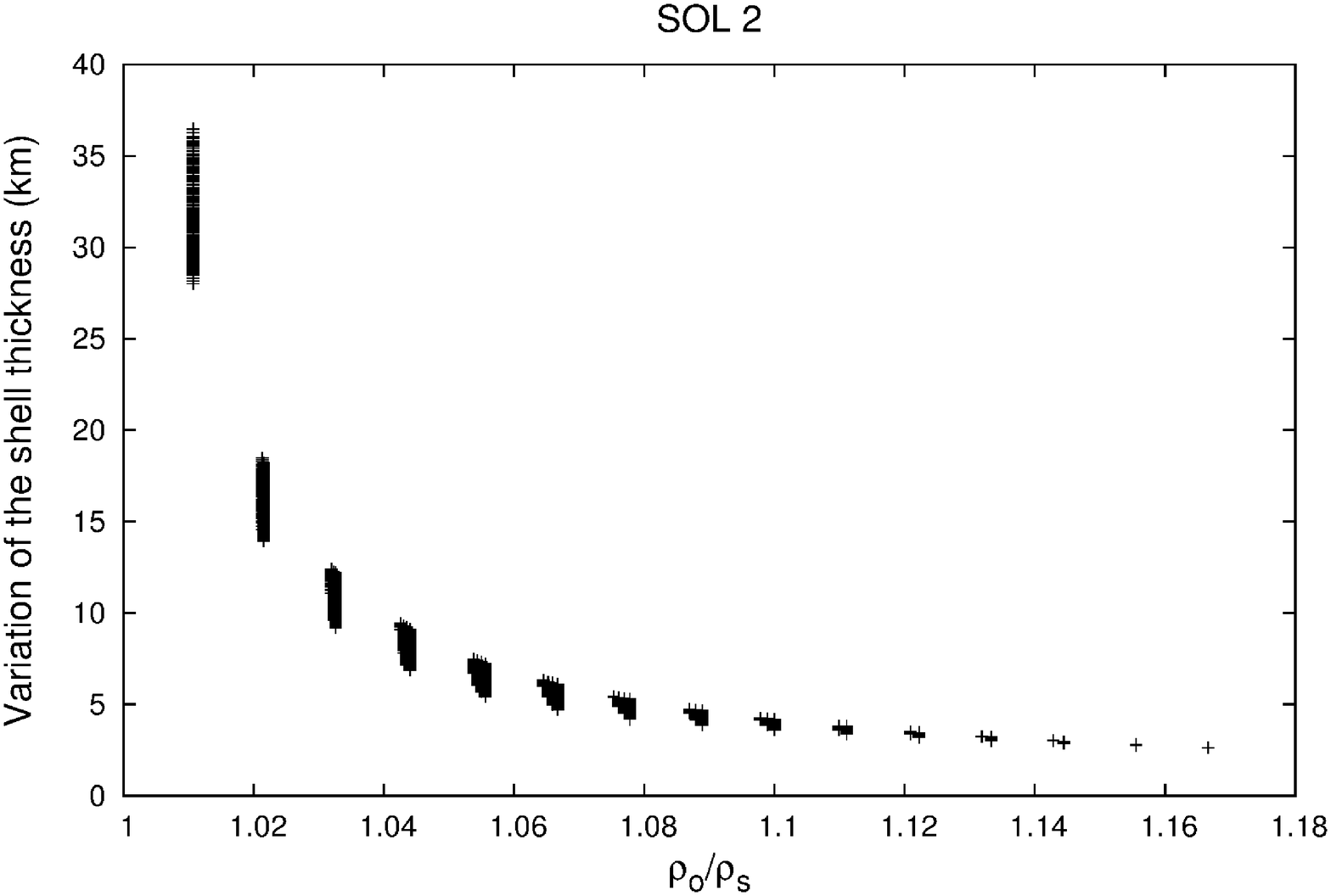}
\end{tabular}
\caption[Thickness variations of the shell of our Titans.]{Thickness variations of the shell of our Titans. Here only the $1\sigma$ solutions are displayed, the results being not significantly 
different when extended to $3\sigma$.\label{fig:varthick}}
\end{figure}

 Fig.~\ref{fig:iso2sol} shows some properties of our model Titans. We can see in particular that the gravity field constrains the degree-two compensation 
factor $f$ to a range roughly 0.75-0.95, very similar to the results obtained by \citet{hnzi2013}. In other words, 
the bottom load is mostly but not entirely compensated. The differences between the maximum and minimum thickness of the shell are shown in Fig.\ref{fig:varthick}. 
They can reach 38~km when the densities of the core and of the shell are very close, while they do not exceed 60 meters when Titan is in hydrostatic equilibrium.

We note that our approach assumes that all the non-hydrostatic effects are contained within the ice shell, i.e. the core is hydrostatic. This assumption can be 
justified based on the strong observed correlation at degree-3 between the shell surface topography and the gravity \citep{hnzi2013}; a low-rigidity (and likely hydrostatic) 
core is also indicated by the large tidal response of Titan \citep{ijdslaarrt2012}.


\placefigure{fig:iso2sol}

\placefigure{fig:varthick}

\clearpage

\section{Results\label{sec:results}}

\par We here present the results of the numerical simulations of the rotational dynamics of Titan for 71,046 interior models, 20,526 consistent with SOL1 (2,237 of them at $1\sigma$) 
and 50,500 with SOL2 (2,681 of them at $1\sigma$).

\subsection{Longitudinal librations}

\par
\begin{figure}[ht]
\centering
\begin{tabular}{cc}
  \includegraphics[width=0.45\textwidth]{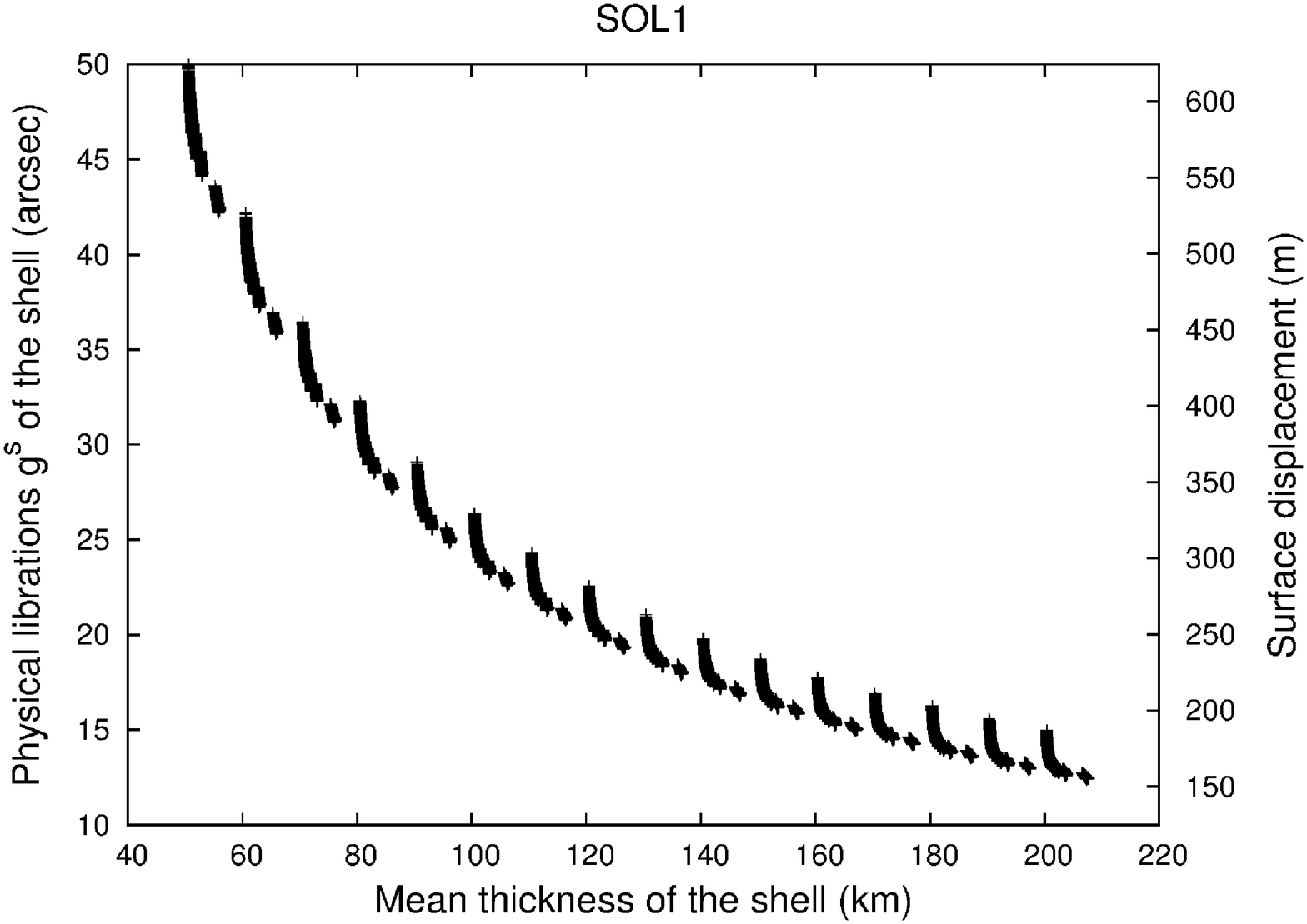} & \includegraphics[width=0.45\textwidth]{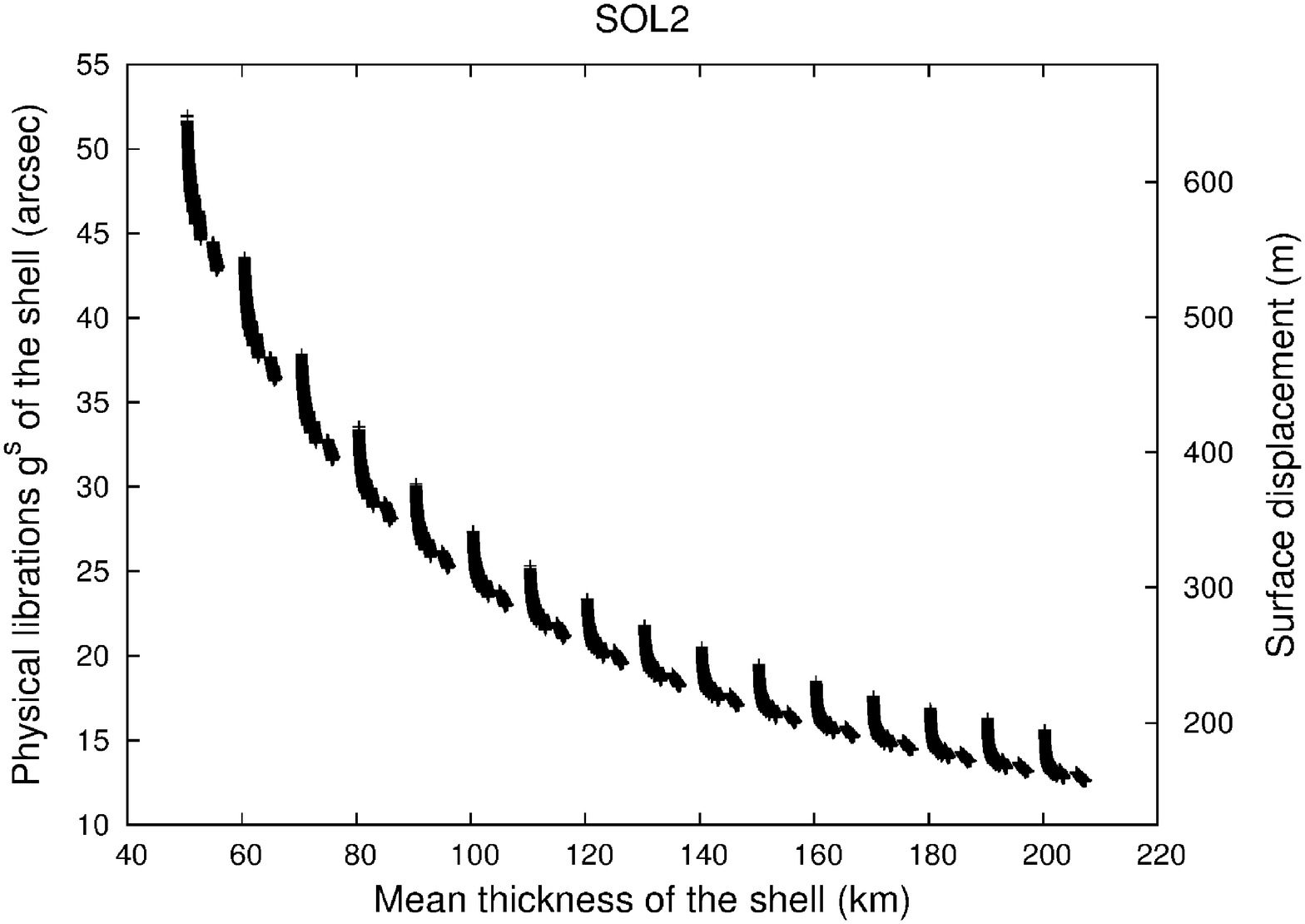}
\end{tabular}
\caption{Longitudinal librations of the shell.\label{fig:libshell}}
\end{figure}

 We here show the amplitude of the physical longitudinal librations $\gamma^{s,c}$ at the orbital frequency. This amplitude is denoted $g^s$ for the shell (Fig.\ref{fig:libshell}) 
and $g^c$ for the core. These amplitudes have been obtained numerically, but they are in very good agreement with the ones given by the analytical formulae 
(see Appendix \ref{sec:analibr}).

\placefigure{fig:libshell}

\par We get an amplitude of libration $g^s$ that can reach 600 meters for a thin shell. One important caveat here is that we are neglecting the rigidity of the ice 
shell, which tends to oppose librational motion \citep{gm2010}, and which in the case of Titan is thought to be large \citep{hnzi2013}. When elasticity is considered, 
the amplitude of libration could be 10 times smaller \citep{vbt2013,rrc2014}; \citep{jv2014} have found a similar result for Europa. 


\par \citet{vrkb2009} have shown that when the atmospheric torque was considered, then the main librations were the semi-annual, with a period of 14.29 years. 
That study used the atmospheric torque published by \citep{tn2005}. \citep{rrc2014} have shown that using the global climate model of \citep{lbrc2012} gives a smaller amplitude.
We here do not consider the atmospheric torque and do not detect a significant 14.29-y periodic oscillation.



\par For the core, we predict a much smaller amplitude of libration, between 2.1 and 2.6 arcmin. The two periods of free librations associated are respectively
between 200 and 400 days, and between 2.5 and 2.7 years. These periods do not allow any resonance with forcing frequencies, and so the libration amplitudes are not raised.






\subsection{A resonant obliquity of the shell}

\par The obliquity of the shell is the only rotational quantity that has been measured, beside the spin rate. Its value of $0.31^{\circ}\pm0.05^{\circ}$ \citep{mi2012}
is surprisingly high (see Sec.\ref{sec:intro}). To the best of our knowledge, the only satisfying explanation present in the literature \citep{bvyk2011} is a resonance 
between the free librations of the obliquity and a forced oscillation due to either the Solar gravitational perturbation of 29.46 years or the regression of Titan's 
ascending node around Saturn, the period associated being 703.51 years. A resonance is usually a strong phenomenon, that is efficient over a limited range of parameters.
An issue is the probability that Titan be affected by such a resonance.

\begin{figure}[ht]
\centering
\begin{tabular}{cc}
  \includegraphics[width=0.45\textwidth]{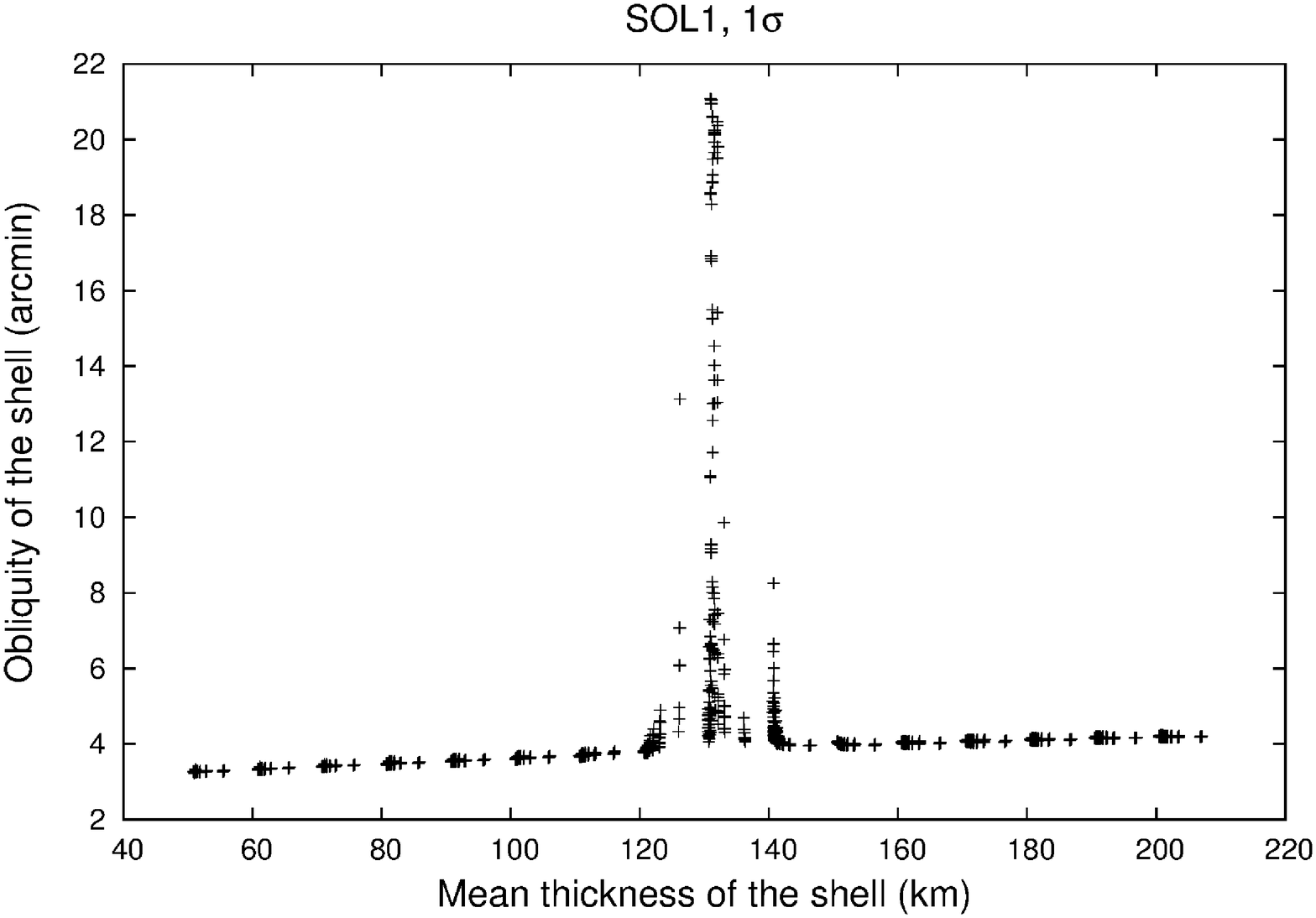} & \includegraphics[width=0.45\textwidth]{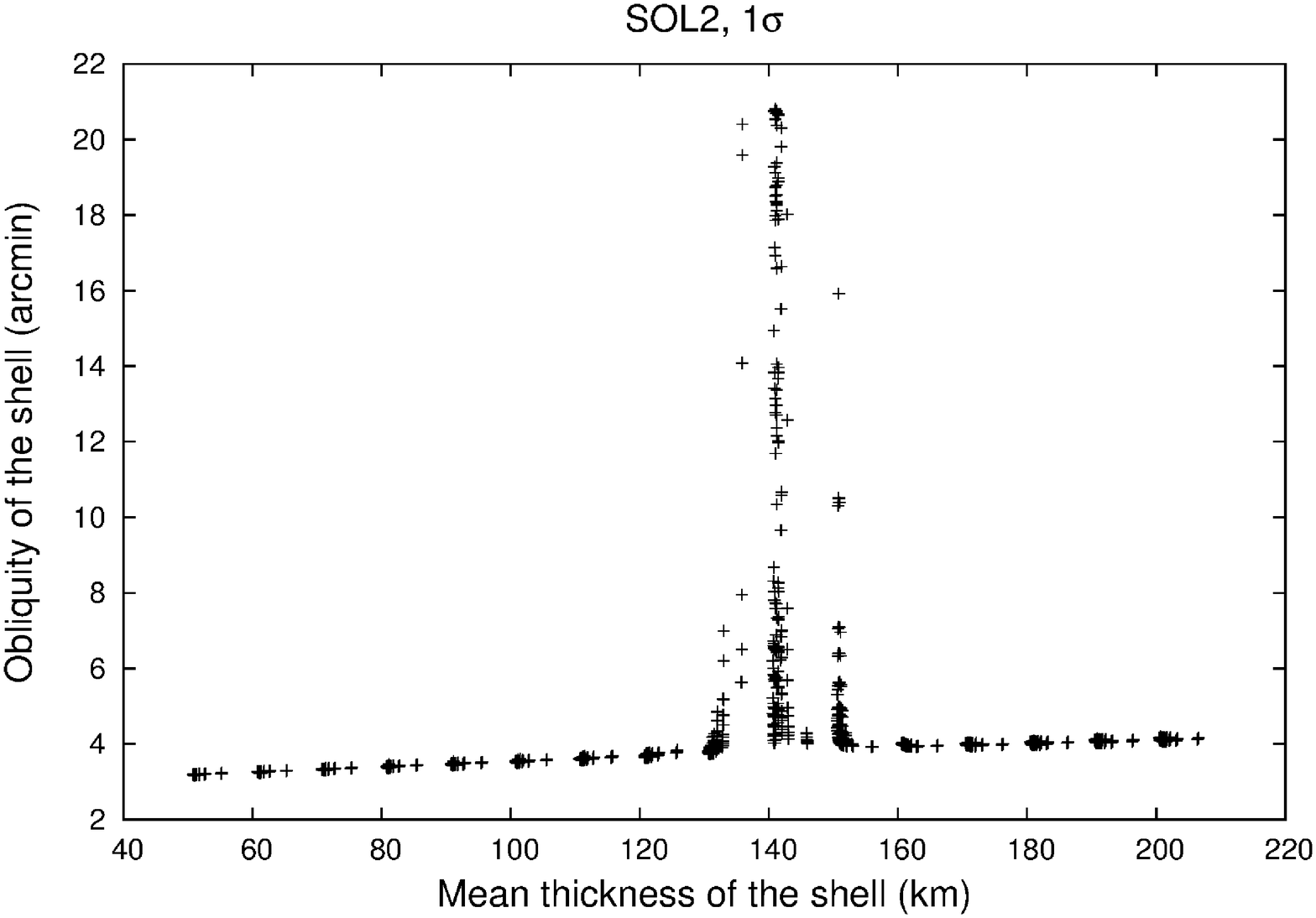} \\
  \includegraphics[width=0.45\textwidth]{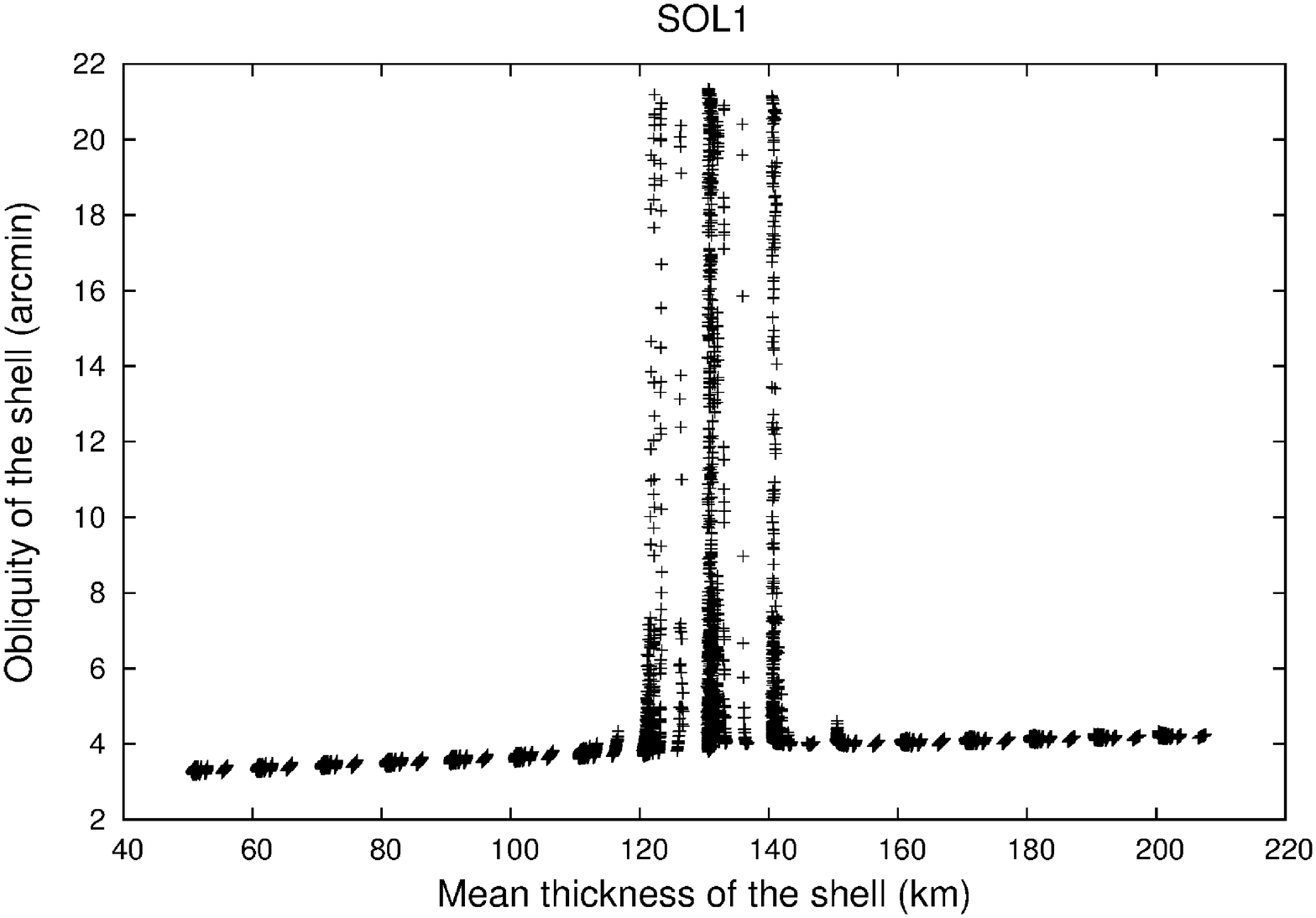} & \includegraphics[width=0.45\textwidth]{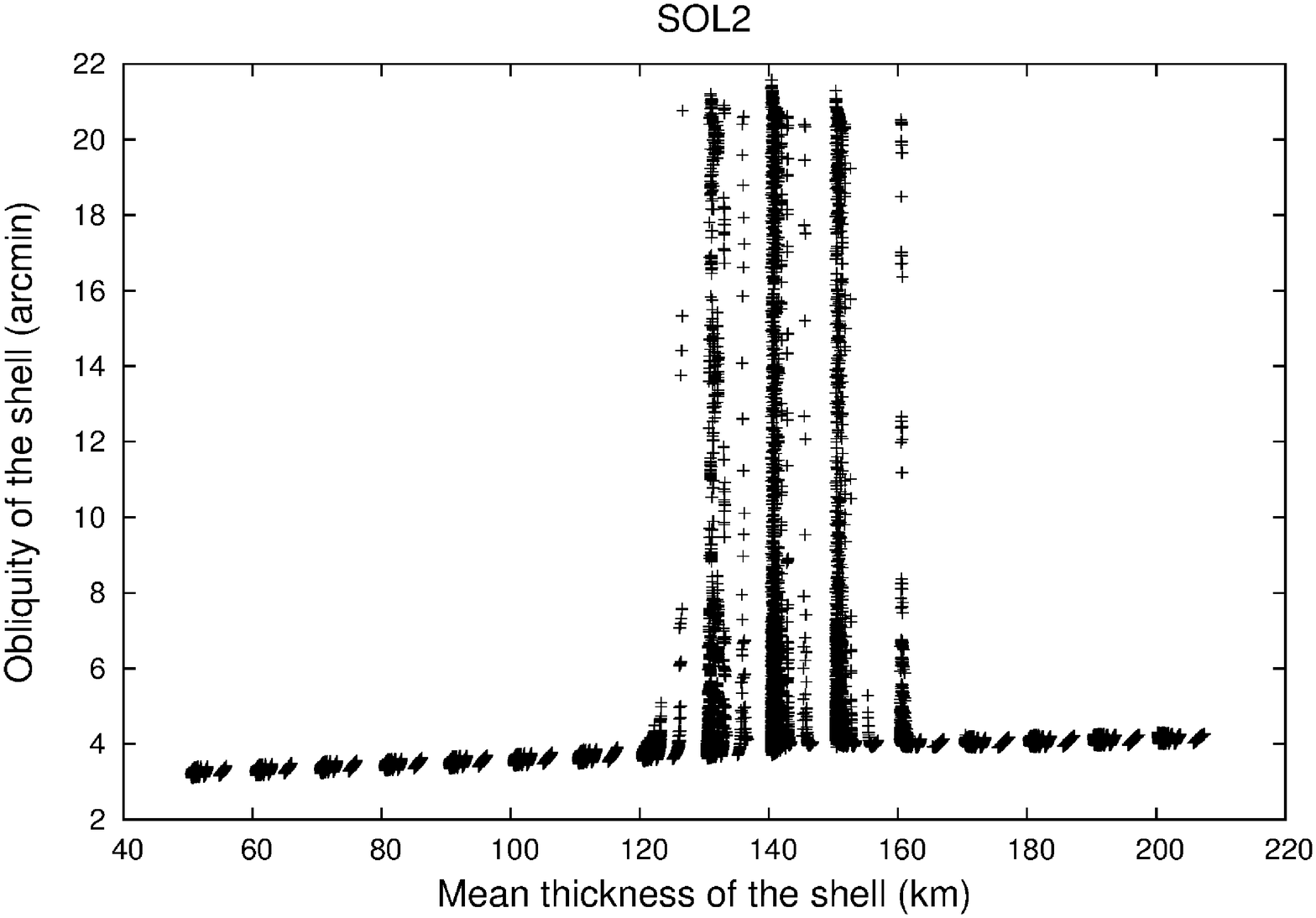}
\end{tabular}
\caption{Influence of the thickness of the shell on its obliquity.\label{fig:oblishtsh}}
\end{figure}

\begin{figure}[ht]
\centering
\begin{tabular}{cc}
  \includegraphics[width=0.45\textwidth]{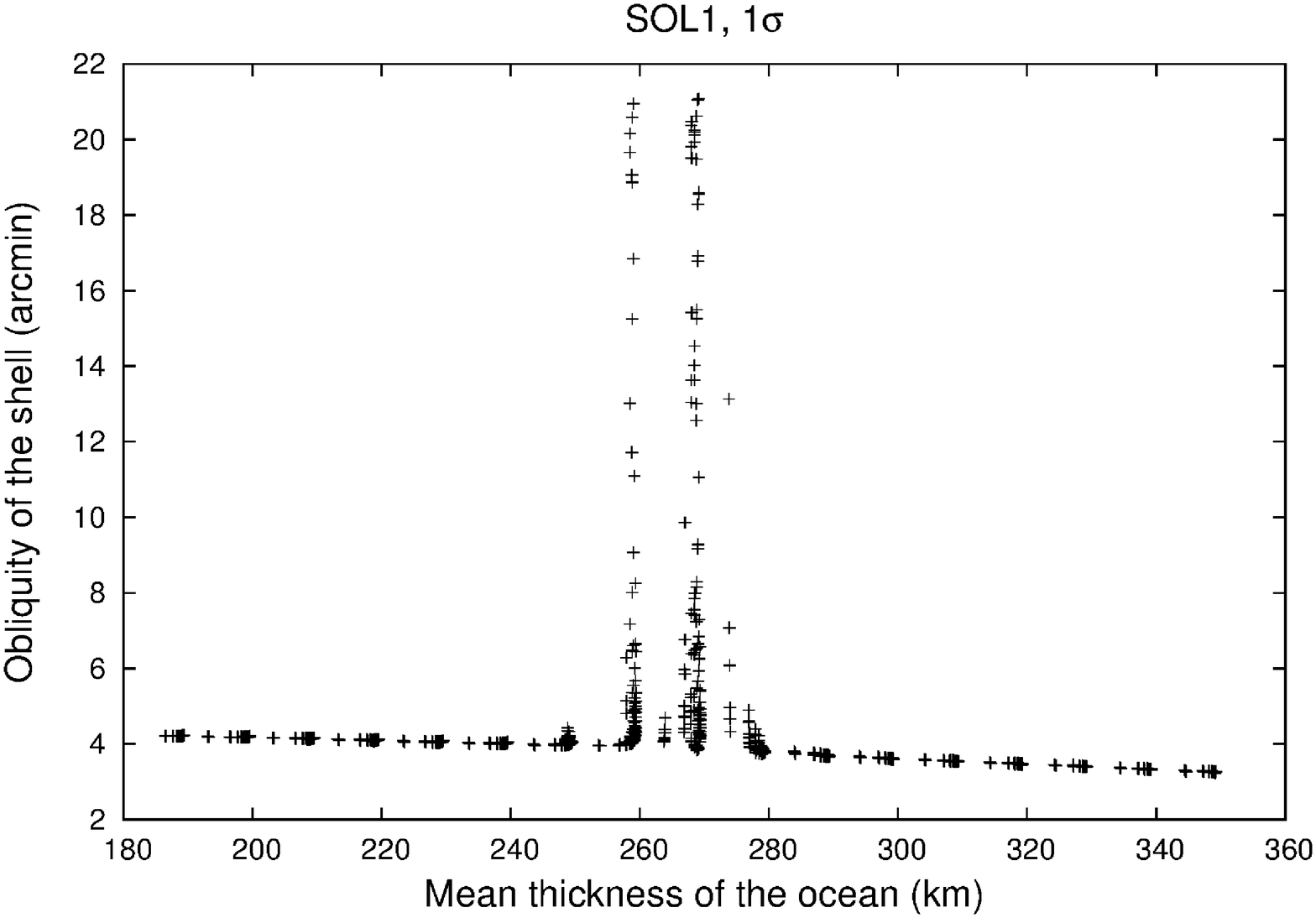} & \includegraphics[width=0.45\textwidth]{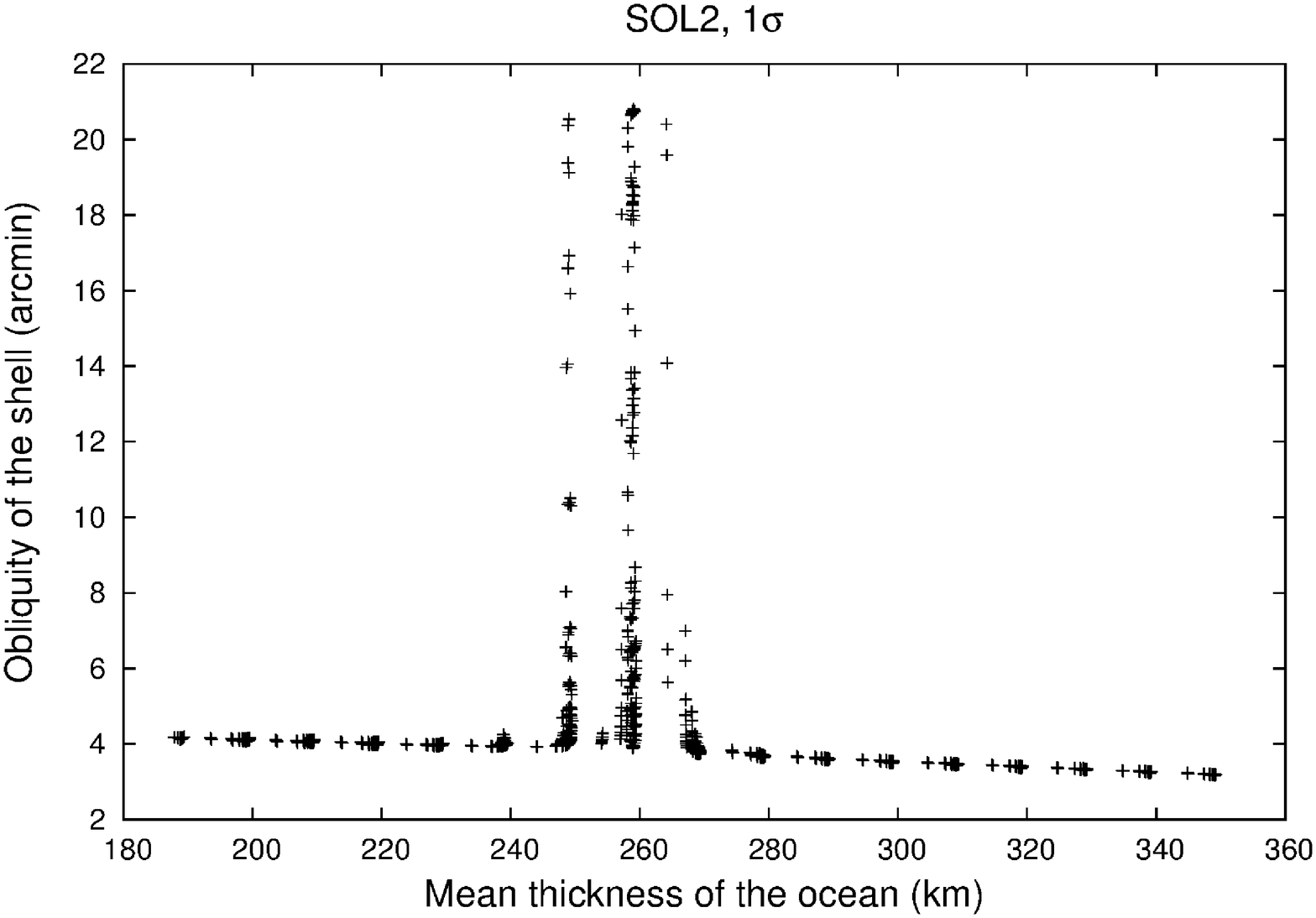} \\
  \includegraphics[width=0.45\textwidth]{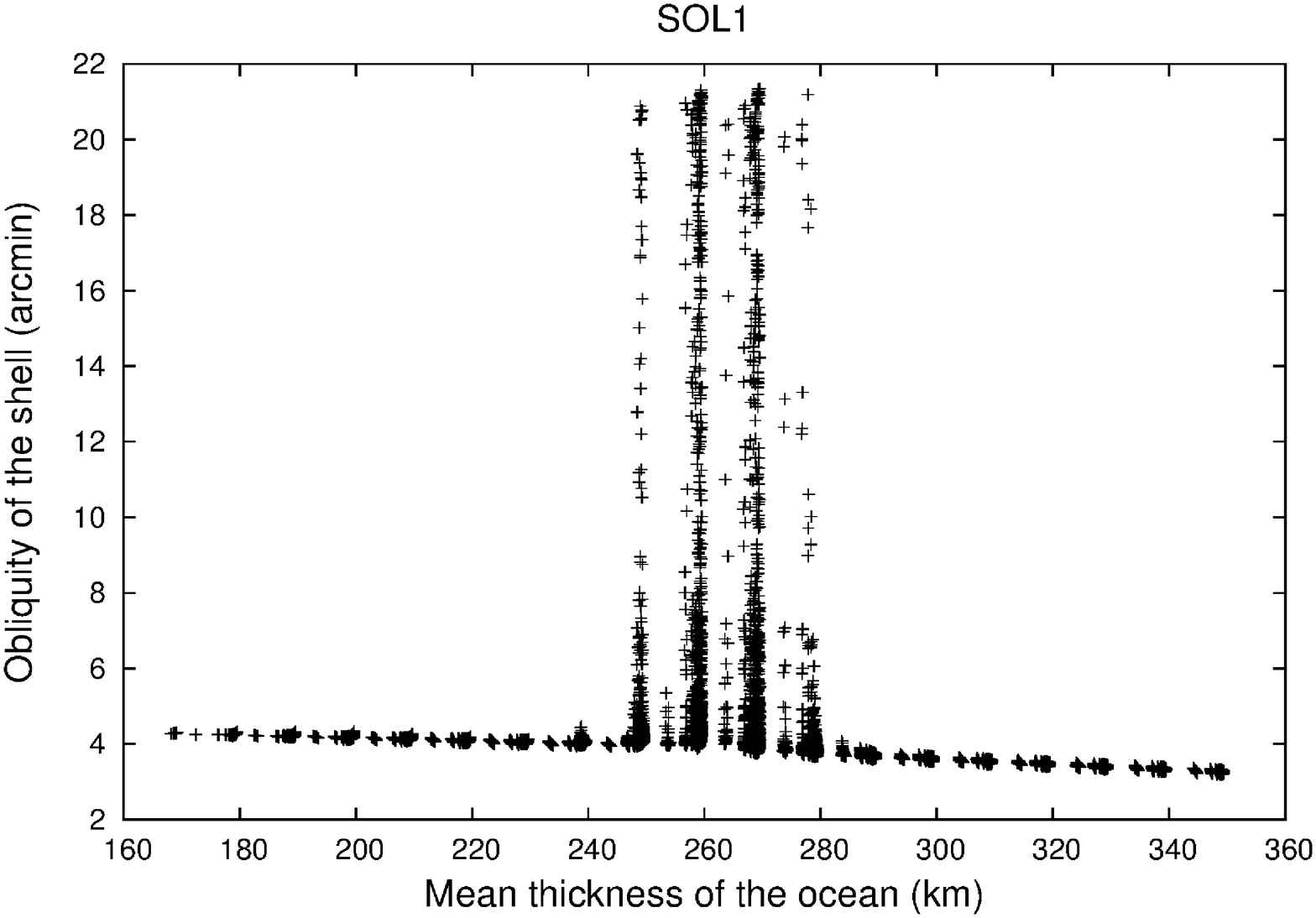} & \includegraphics[width=0.45\textwidth]{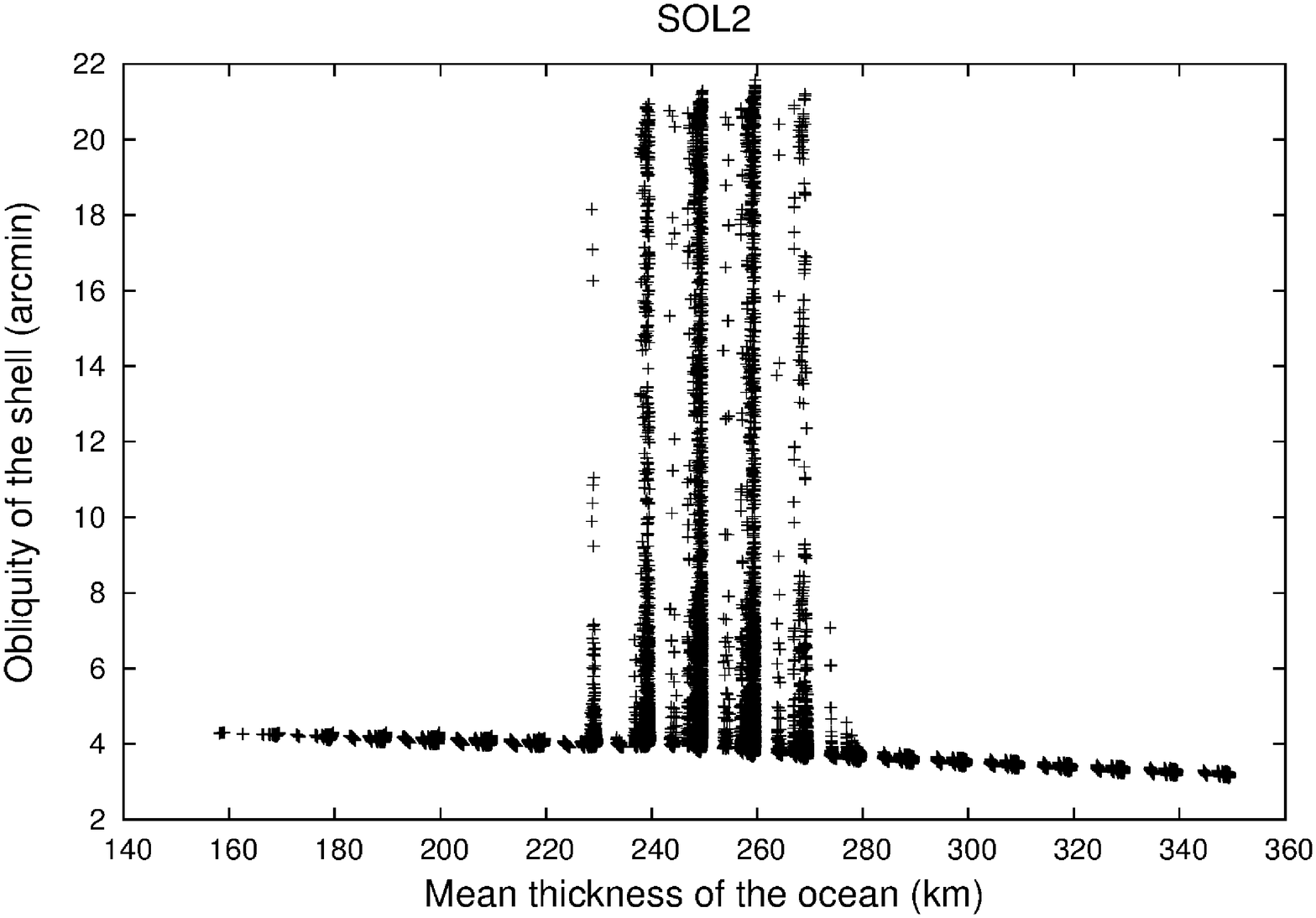}
\end{tabular}
\caption{Influence of the thickness of the ocean on the obliquity of the shell.\label{fig:oblishtoc}}
\end{figure}

\placefigure{fig:oblishtsh}
\placefigure{fig:oblishtoc}

\par Figs.\ref{fig:oblishtsh} and \ref{fig:oblishtoc} show the mean obliquity of the shell with respect to the mean thickness of the shell and the ocean. We plot here 
the results for model Titans with a gravity field within the $1-\sigma$ limit of the measured one, and within the $3-\sigma$ limit. We can see two regimes for the obliquity. 
Most of the solutions lie on a nearly constant line, indicating an obliquity between 3 and 4.2 arcmin. This is the obliquity given by the classical theory of a rigid body
in a secular regime. These numbers are too small with respect to the measured one, i.e. $\approx18$~arcmin. However, another regime with much higher obliquities can be seen, for thicknesses of the shell
and of the ocean of $\approx130$ and $\approx270$~km for SOL1, and $\approx140$ and $\approx260$~km for SOL2. Since a resonance is suspected, we determine the periods of the
free oscillations associated with the obliquity.

\par In adapting a result by \citet{bvyk2011}, the frequencies of these free librations $\sigma_3$ and $\sigma_4$ are:

\begin{eqnarray}
  \sigma_3 & = & -\frac{Z+\sqrt{\Delta}}{2C^cC^s}, \label{eq:sigma3} \\
  \sigma_4 & = & -\frac{Z-\sqrt{\Delta}}{2C^cC^s}, \label{eq:sigma4}
\end{eqnarray}
with

\begin{eqnarray}
  Z        & = & K(C^c+C^s)-C^s\kappa^c-C^c\kappa^s,                                        \label{eq:Z} \\
  \Delta   & = & -4C^cC^s\left(\kappa^c\kappa^s-K\left(\kappa^c+\kappa^s\right)\right)+Z^2, \label{eq:Delta} \\
  K        & = & \frac{3\beta^o}{n_6}\left(C^c-A^c+C_b^o-A_b^o\right),                      \label{eq:K} \\
  \kappa^s & = & \frac{3n_6}{2}\left(C^s-A^s+C_t^o-A_t^o\right),                              \label{eq:kappas} \\
  \kappa^c & = & \frac{3n_6}{2}\left(C^c-A^c+C_b^o-A_b^o\right).                              \label{eq:kappac}
\end{eqnarray}

\par
\begin{figure}[ht]
\centering
\begin{tabular}{cc}
  \includegraphics[width=0.45\textwidth]{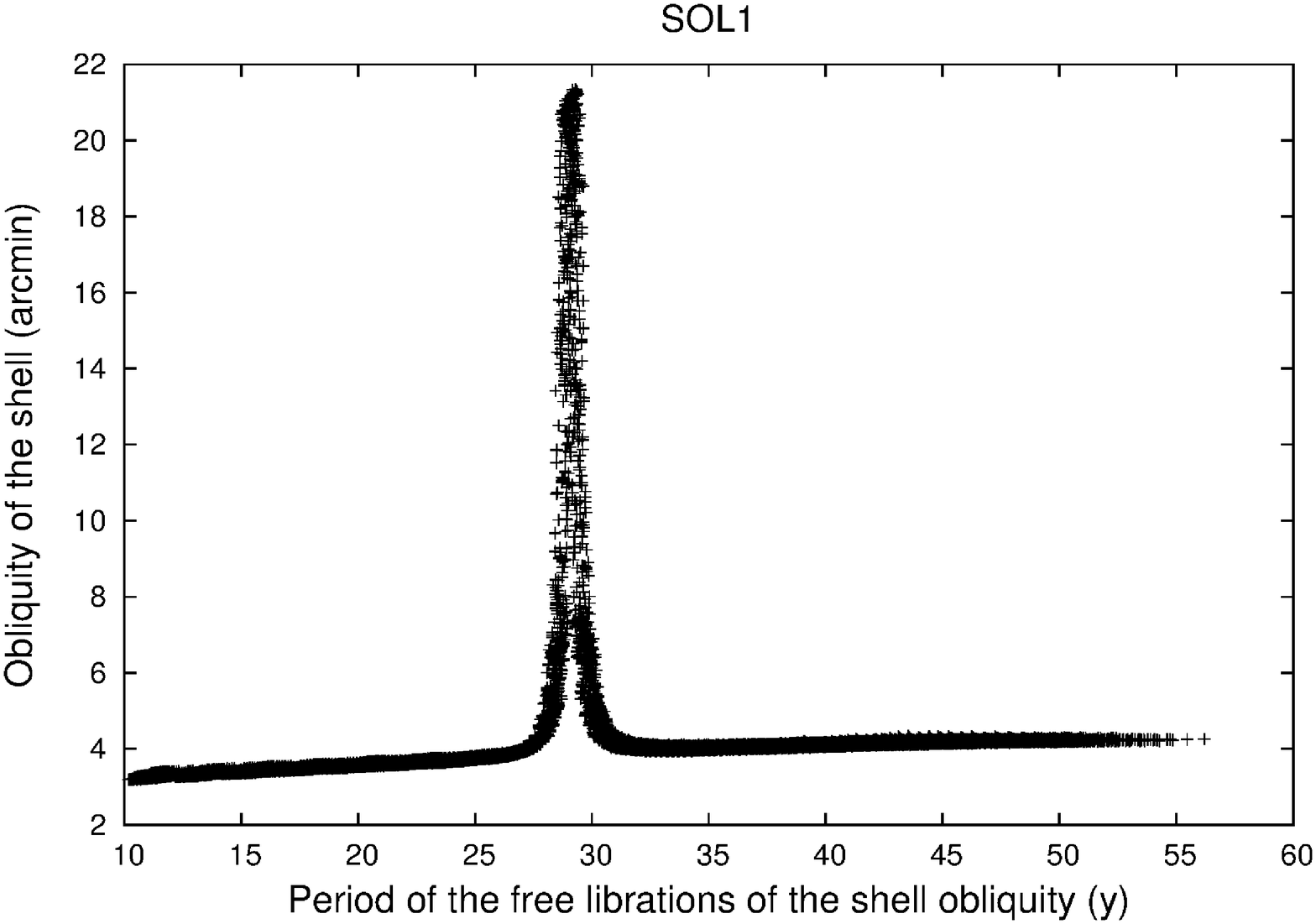} & \includegraphics[width=0.45\textwidth]{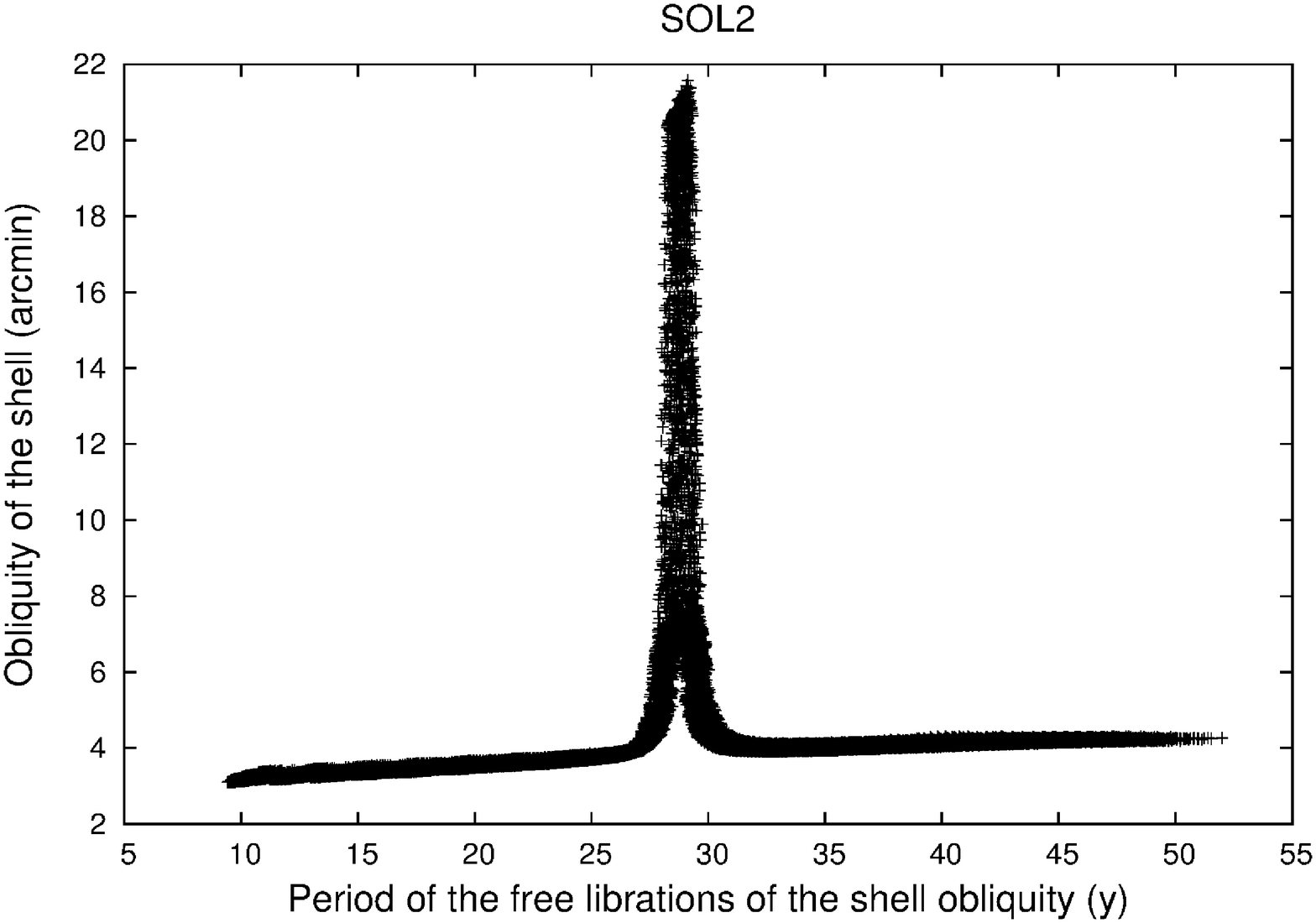}
\end{tabular}
\caption[The resonance raising the obliquity of Titan.]{The resonance raising the obliquity of Titan. In the resonance, the averaging lacks 
reliability since the free oscillations are tricky to remove. The information to keep in mind is that in the resonance, the obliquity of the shell 
is likely to be much bigger than suggested by the secular behavior out of the resonance.\label{fig:resonance}}
\end{figure}

 The periods associated, i.e. $T_3=2\pi/\sigma_3$ and $T_4=2\pi/\sigma_4$, are respectively between 200 and 260 years, and between 10 and 55 years. The range 200-260 years does not 
correspond to any obvious forcing, while the annual forcing, 29.46-yr periodic, can resonate with $\sigma_4$. Figure \ref{fig:resonance} shows the mean obliquity vs. $T_4$.

\placefigure{fig:resonance}

\par The correlation between the obliquity and the resonance with the 29.46-yr periodic annual forcing is obvious. From the value of $T_4$, we estimate that 2,202 model 
Titans out of 20,546 for SOL1, i.e. $10.73\%$, are affected, and 6,540 out of 50,500 for SOL2, i.e. $12.95\%$, within the $3-\sigma$ limit. If we restrict to the  $1-\sigma$ limit,
then we have 234 model Titans over 2,237 for SOL1, i.e. $10.46\%$, and 354 over 2,665 for SOL2, i.e. $13.28\%$.

\par When a trajectory is resonant, getting rid of the free librations is very tricky and nearly impossible when working on a whole set of trajectories as we do in this Section. 
So, the numbers for the mean obliquity of the shell given in Fig.\ref{fig:oblishtsh} to \ref{fig:resonance} should not be considered as accurate in the resonant regime. The 
take-home message is that the resonance can raise the obliquity of these Titans to numbers bigger than the measured value. We will simulate some individual trajectories
in Section~\ref{sec:trajectories}.

\subsection{Polar motion}

\par We check the amplitude of the polar motion, often supposed to be small. Actually it is, the maximum amplitude being 3.5 km for the shell and 4.5 km for the core.
Most of the trajectories have a polar motion much smaller than that.




\section{Simulating the obliquity of the real Titan\label{sec:trajectories}}

\par The resonant obliquity of Titan's shell is highly sensitive to one frequency of the free oscillations of the obliquity, $\sigma_4$, itself dependent on the interior structure. For this reason, 
we propose here to build a synthetic representation of the obliquity of the shell with respect to the relevant frequency. This obliquity is given as a sum of a time series 
composed of trigonometric terms with numerical amplitudes. Using a synthetic theory to express the obliquity of a celestial body has already been done for Mercury \citep{nd2012,nl2013}.
Here the strategy is different since the periodic contributions involved, due to the orbital motion of Saturn about the Sun, have a much shorter period ($\approx30$ years) than the
relevant ones for Mercury, due to the regressional motion of Mercury's ascending node, the associated period being $\approx250$ kyears.

\subsection{A synthetic model}

\par Our goal is to express the obliquity of the shell as a sum:

\begin{equation}
  \label{eq:Ks0}
  K^s(t) = A_0+\sum_i A_i \cos\phi_i(t),
\end{equation}
where $A_i$ are real amplitudes, and $\phi_i(t)$ linear functions of degree one of time. For that, we consider 13 models of Titan (see Tab.\ref{tab:restitans}) that fall into the resonance and
correspond to the gravity solution SOL1 within the 1-sigma limit. Here the key parameter is the frequency $\sigma_4$ and the models have been chosen among this criterion, 
so choosing them in SOL2 instead of SOL1 would not change the results.

\placetable{tab:restitans}

\begin{landscape}
\begin{table}[ht]
  \centering
  \caption[Internal structure of 13 resonant Titans.]{Internal structure of the 13 resonant Titans that we use in this Section, complying with the SOL1 gravity field. 
  The densities are in $kg/m^3$, the radii in km, and $f$ is the compensation factor. $2440$ km have been substracted from the radii of the shell-ocean boundary 
  $a_o$, $b_o$ and $c_o$, and $2170$ km have been substracted from the radii of the core-ocean boundary $a_c$, $b_c$ and $c_c$. The outer radii are the measured 
  ones (Tab.\ref{tab:titanshape}). All models had 130~km mean shell thicknesses, and all except N=58 had 270 km mean ocean thicknesses. The mean ocean thickness
  of the model 58 was 260~km.\label{tab:restitans}}
  \begin{tabular}{rrrrrrrrrrrrrc}
   N & $\rho_s$   & $\rho_o$   & $\rho_c$    & $a_o$    & $b_o$    & $c_o$     & $a_c$     & $b_c$     & $c_c$     & f           & $J_2$           & $C_{22}$        & Shell thickness\\
     &            &            &             &          &          &           &           &           &           &             & $\times10^{-5}$ & $\times10^{-6}$ &   Min - Max \\
     \hline
  58 &    $900$   & $950$      & $2486.396$ & $2.282$ & $3.517$ &  $8.587$ & $14.983$ & $14.731$ & $14.648$ & $80\%$      &  $3.147$        &  $9.973$        & $125.843$-$132.868$ \\
 791 &    $910$   & $960$      & $2501.093$ & $2.303$ & $3.538$ &  $8.544$ &  $4.981$ &  $4.732$ &  $4.649$ & $82\%$      &  $3.176$        &  $9.949$        & $125.885$-$132.847$ \\
1228 &    $920$   & $950$      & $2502.910$ & $0.469$ & $2.733$ & $11.184$ &  $4.980$ &  $4.732$ &  $4.649$ & $82.5\%$    &  $3.183$        &  $9.951$        & $123.246$-$134.681$ \\
1230 &    $920$   & $950$      & $2502.910$ & $0.523$ & $2.757$ & $11.105$ &  $4.980$ &  $4.732$ &  $4.649$ & $83.5\%$    &  $3.207$        &  $9.966$        & $123.325$-$134.627$ \\
1231 &    $920$   & $950$      & $2502.910$ & $0.550$ & $2.768$ & $11.067$ &  $4.980$ &  $4.732$ &  $4.649$ & $84\%$      &  $3.219$        &  $9.974$        & $123.363$-$134.600$ \\
1333 &    $920$   & $960$      & $2498.704$ & $1.574$ & $3.211$ &  $9.600$ &  $4.981$ &  $4.731$ &  $4.648$ & $81.5\%$    &  $3.168$        &  $9.961$        & $124.830$-$133.576$ \\
1336 &    $920$   & $960$      & $2498.704$ & $1.636$ & $3.239$ &  $9.511$ &  $4.981$ &  $4.731$ &  $4.648$ & $83\%$      &  $3.204$        &  $9.984$        & $124.919$-$133.514$ \\
1485 &    $920$   & $980$      & $2490.291$ & $2.692$ & $3.695$ &  $7.998$ &  $4.981$ &  $4.731$ &  $4.648$ & $80\%$      &  $3.148$        &  $9.988$        & $126.432$-$132.457$ \\
1487 &    $920$   & $980$      & $2490.291$ & $2.721$ & $3.708$ &  $7.956$ &  $4.981$ &  $4.731$ &  $4.648$ & $81\%$      &  $3.173$        & $10.004$        & $126.474$-$132.429$ \\
1488 &    $920$   & $980$      & $2490.291$ & $2.735$ & $3.714$ &  $7.936$ &  $4.981$ &  $4.731$ &  $4.648$ & $81.5\%$    &  $3.186$        & $10.012$        & $126.494$-$132.415$ \\
1783 &    $930$   & $960$      & $2496.315$ & $0.442$ & $2.705$ & $11.239$ &  $4.981$ &  $4.731$ &  $4.648$ & $82.5\%$    &  $3.197$        &  $9.996$        & $123.191$-$134.708$ \\
1858 &    $930$   & $970$      & $2492.109$ & $1.554$ & $3.190$ &  $9.642$ &  $4.981$ &  $4.731$ &  $4.648$ & $81.5\%$    &  $3.181$        & $10.005$        & $124.788$-$133.596$ \\
1860 &    $930$   & $970$      & $2492.109$ & $1.596$ & $3.209$ &  $9.581$ &  $4.981$ &  $4.731$ &  $4.648$ & $82.5\%$    &  $3.206$        & $10.021$        & $124.849$-$133.554$ \\
\hline
  \end{tabular}
\end{table}
\end{landscape}

\par For each of these 13 models, we simulate the rotation corresponding to the dynamical equilibrium. For that, we use the algorithm of \citep{ndc2014} already mentionned,
consisting in removing iteratively the free oscillations from the initial conditions, after frequency analysis. The difficulty here comes from the quasi-resonant condition,
because the period associated with $\sigma_4$ is close to the period of the annual forcing, i.e. $29.45716$ years. To bypass this problem, we must integrate over a 
large time interval so that the distance between the frequencies gets bigger than twice the frequency whose period is the integration interval. In practice, we have integrated
over 36,800 years. The TASS1.6 ephemeris are valid over 9,000 years, but since they are composed of trigonometric series, they can be extrapolated without diverging.

\par After obtaining the trajectories and frequency analysis of the obliquities, we get:

\begin{equation}
  \label{eq:Ksserie}
  K^s(t,\sigma_4) \approx A_0(\sigma_4)+A_1(\sigma_4)\cos\phi_1(t)+A_2(\sigma_4)\cos\phi_2(t),
\end{equation}
with

\begin{eqnarray}
  \phi_1(t) & = & 0.20436788 \, t - 1.08611, \label{eq:phi1} \\
  \phi_2(t) & = & 0.63989736 \, t + 1.22080, \label{eq:phi2}
\end{eqnarray}
the time origin being J2000, the time in years, and the angles in radians. The periods of these forced oscillations are respectively $30.74475$ and $9.81906$ years. 
$\phi_1$ and $\phi_2$ can be reconstructed from the elements of Tab.\ref{tab:inclitass}, numbered from (1) to (4). We have $\phi_1=(1)-(4)$, and $\phi_2=(3)-(4)$.

\par Tab.\ref{tab:synth} gathers the coefficients $A_i$ given by the frequency analysis. The error due to this representation is always smaller than $1.1$ arcmin over 100 years.
This is actually a maximum of the difference between the obliquity given by the formula (\ref{eq:Ksserie}) and the one resulting from our numerical simulation. This difference
can be due to neglected oscillating contributions in the Eq.(\ref{eq:Ksserie}), but also to a residual of free oscillation in the numerical simulation, that noises the signal.

\placetable{tab:synth}

\begin{table}[ht]
  \centering
  \caption[Synthetic representation of the obliquity of Titan.]{Synthetic representation of the obliquity of the shell. The error 
  due to the synthetic representation is estimated over 100 years from J2000. $T_4$ is the period associated with $\sigma_4$ measured
  by frequency analysis, while $T_4^{\star}$ is given by the Eq.(\ref{eq:sigma4}). The last column gives the computed obliquity 
  at the observation date, i.e. $\approx$2007.2. In this table, the models 58 and 1487 are the closest to the measured obliquity.\label{tab:synth}}
  \begin{tabular}{rrrrrrrrr}
  \hline
      N & $A_0$      & $A_1$      & $A_2$      & Error    & $T_4$      & $T_4^{\star}$ & $\Delta T_4$ & @J2007.2 \\
        & (arcmin)   & (arcmin)   & (arcmin)   & (arcmin) & (years)    & (years)         &              & (arcmin) \\
  \hline
     58 & $18.46293$ & $-3.79374$ &  $0.58755$ & $0.6$    & $29.59253$ & $29.25391$      & $1.14\%$     & $15.4751$ \\
    791 & $11.12233$ &  $3.70234$ & $-0.57339$ & $0.8$    & $29.23168$ & $28.75465$      & $1.63\%$     & $14.0382$ \\
   1228 &  $9.10463$ & $-3.67034$ &  $0.56454$ & $0.9$    & $29.74431$ & $29.26631$      & $1.61\%$     &  $6.2105$ \\
   1230 & $42.51503$ &  $3.75107$ & $-0.59207$ & $0.3$    & $29.40183$ & $28.85206$      & $1.87\%$     & $45.4592$ \\
   1231 & $11.05850$ &  $3.68666$ & $-0.57384$ & $0.8$    & $29.23102$ & $28.65278$      & $1.98\%$     & $13.9594$ \\
   1333 &  $9.48745$ & $-3.68069$ &  $0.56667$ & $1.1$    & $29.73195$ & $29.30645$      & $1.43\%$     &  $6.5856$ \\
   1336 & $10.17817$ &  $3.67598$ & $-0.56950$ & $0.8$    & $29.20998$ & $28.68037$      & $1.81\%$     & $13.0731$ \\
   1485 & $15.61029$ & $-3.74199$ &  $0.58342$ & $1.0$    & $29.61870$ & $29.26416$      & $1.20\%$     & $12.6667$ \\
   1487 & $12.79944$ &  $3.71361$ & $-0.57774$ & $0.7$    & $29.26101$ & $28.83270$      & $1.46\%$     & $15.7218$ \\
   1488 &  $7.10257$ &  $3.58116$ & $-0.53779$ & $0.8$    & $29.08751$ & $28.62520$      & $1.59\%$     &  $9.9381$ \\
   1783 &  $9.48544$ & $-3.66939$ &  $0.56698$ & $1.1$    & $29.73153$ & $29.25060$      & $1.62\%$     &  $6.5944$ \\
   1858 &  $9.77734$ & $-3.67887$ &  $0.56840$ & $0.9$    & $29.72422$ & $29.29491$      & $1.44\%$     &  $6.8788$ \\
   1860 & $29.96008$ &  $3.73986$ & $-0.59003$ & $0.4$    & $29.37617$ & $28.87354$      & $1.70\%$     & $32.8957$ \\
   \hline
   \end{tabular}
\end{table}

\par We also check in this table the accuracy of the analytical estimation of $T_4$ by the Eq.(\ref{eq:sigma4}). We can see that this formula slightly underestimates the period. 
Its accuracy, better than $2\%$, is good enough to detect the influence of the resonance, but might be insufficient to predict the resulting obliquity, since this quantity
is highly sensitive to the distance to the exact resonance with the annual forcing. The last column gives $K^s(t=J2007.2)$ calculated from the Eq.(\ref{eq:Ksserie}), this date
is very close to the one of the observation, i.e. March $11^{th}$, 2007.

\par A least-square fit of the obtained numbers for $A_0$, $A_1$ and $A_2$ allows us to write

\begin{eqnarray}
  A_0 & = & \frac{\alpha}{\left|\sigma_4-\sigma\right|}, \label{eq:A0} \\
  A_1 & = & \beta \sgn(\sigma_4-\sigma), \label{eq:A1} \\
  A_2 & = & \gamma \sgn(\sigma-\sigma_4), \label{eq:A2}
\end{eqnarray}
with

\begin{eqnarray}
  \alpha & = & (0.0176084 \pm 0.0002901)\, \textrm{arcmin}, \label{eq:alpha} \\
  \beta  & = & (3.69780 \pm 0.01442)\, \textrm{arcmin}, \label{eq:beta} \\
  \gamma & = & (0.573225 \pm 0.003937)\, \textrm{arcmin}, \label{eq:gamma}
\end{eqnarray}
$\sigma=2\pi/29.45716=0.2132991\,rad/y$ being the frequency of the annual forcing.

\placefigure{fig:A0}

\begin{table}[ht]
  \centering
  \includegraphics[width=0.8\textwidth]{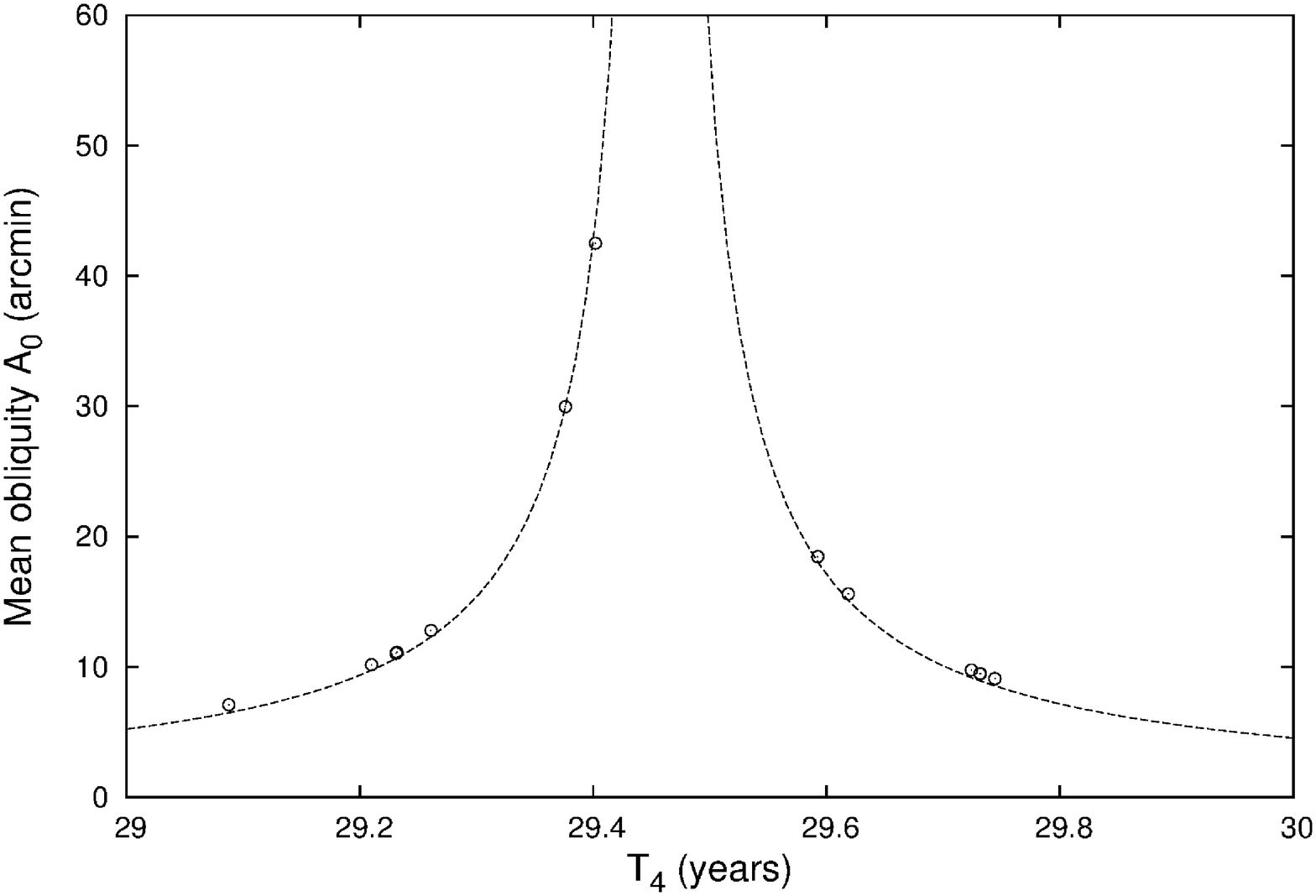}
  \caption[Mean obliquity of Titan.]{Mean obliquity $A_0$ of Titan, fitted with the Eq.(\ref{eq:A0}). The circles correspond to the models given in the Tab.\ref{tab:synth}.\label{fig:A0}}
\end{table}

\par The quantity $A_0$ can be seen as the mean obliquity of Titan, or of its shell, averaged over a long enough interval, here nearly 40 kyears. It should not
be confused with the observed obliquity that is an instantaneous quantity. $A_0$ is plotted in the Fig.\ref{fig:A0}. We can see an asymptotic behavior at the exact resonance,
i.e. $\sigma_4=\sigma$, or $T_4=29.45716$ years. The other two parameters $A_1$ and $A_2$ have a different sign whether $\sigma_4$ is bigger or smaller than $\sigma$.

\subsection{2 solutions}

\par From the Eq.(\ref{eq:Ksserie}) we calculate the obliquity at J2007.2, to be compared to the $(18.6\pm3)$ arcmin measured by \citep{mi2012}. We 
use our Eq.(\ref{eq:Ksserie}) instead of our numerical simulations, to remove the error due to the free oscillations. We get from a linear least squares fit:

\begin{equation}
  \label{eq:obli20072}
  K^s(t=J2007.2,\sigma_4) = \frac{\alpha'}{\|\sigma_4-\sigma\|}+\beta'\sgn(\sigma-\sigma_4),
\end{equation}
with 

\begin{eqnarray}
  \alpha' & = & (0.0176424\pm0.0001689)\,\textrm{arcmin}, \label{eq:alphap} \\
  \beta'  & = & (-2.73983\pm0.1727)\,\textrm{arcmin}. \label{eq:betap}
\end{eqnarray}

\par The Eq.(\ref{eq:obli20072}) is plotted in the Fig.\ref{fig:obli20072}.

\begin{table}[ht]
  \centering
  \includegraphics[width=0.8\textwidth]{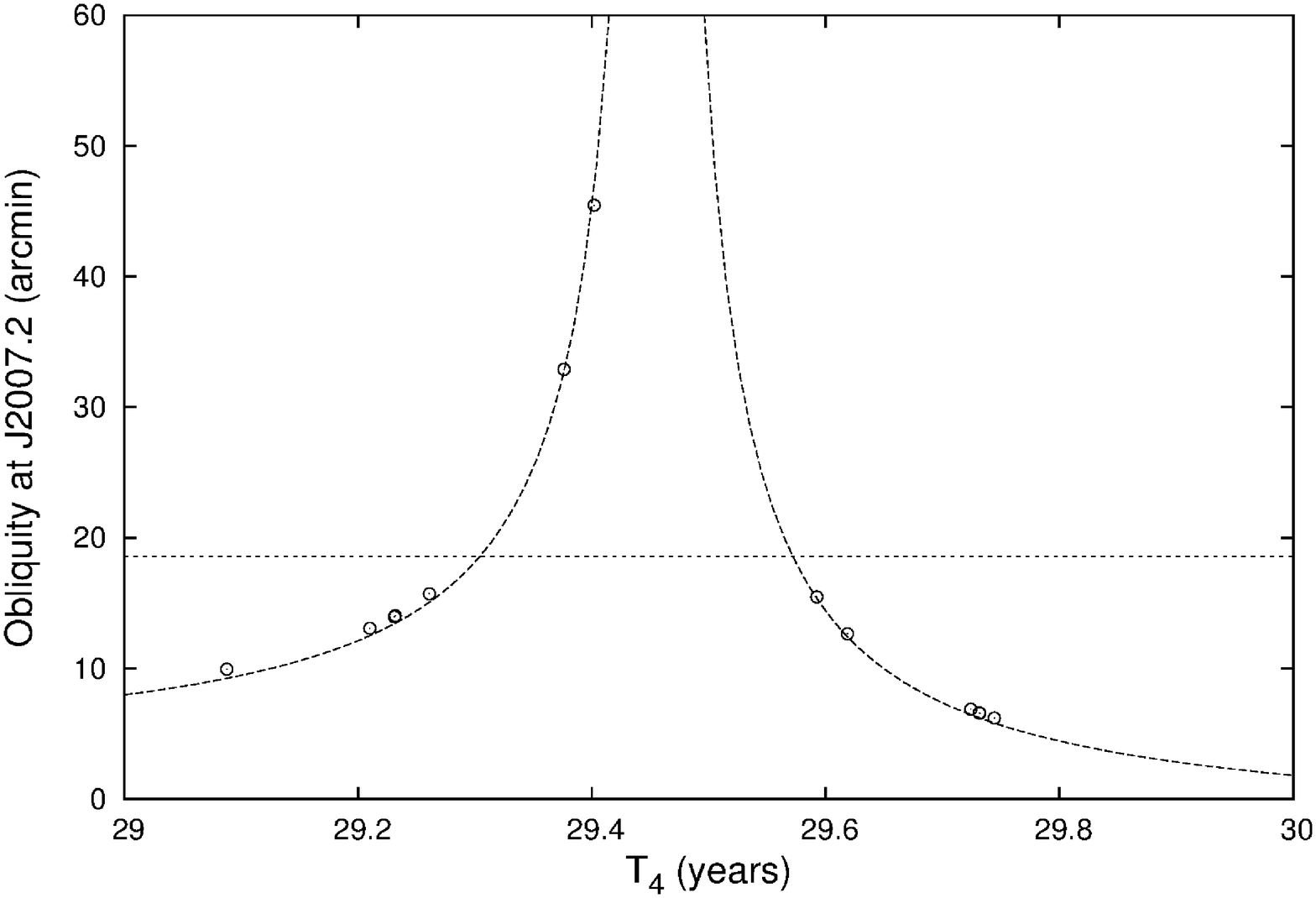}
  \caption[Obliquity of Titan on March 11$^{th}$, 2007.]{Obliquity of Titan on March 11$^{th}$, 2007. The circles correspond to our 13 models, the associated obliquity being calculated
  with the Eq.(\ref{eq:Ksserie}). The plotted function is the Eq.(\ref{eq:obli20072}), while the horizontal line corresponds to the 18.6 arcmin 
  measured by \citep{mi2012}. We can see that the problem has 2 solutions.\label{fig:obli20072}}
\end{table}

\par We can see that 2 frequencies, or periods, can give the measured obliquity of 18.6~arcmin. So, we have what we could call \emph{left solutions},
where $T_4<29.45716$ years, or $\sigma_4>\sigma$, and \emph{right solutions} where $T_4>29.45716$ years, i.e. $\sigma_4<\sigma$.

\begin{figure}[ht]
\centering
\begin{tabular}{cc}
  \includegraphics[width=0.45\textwidth]{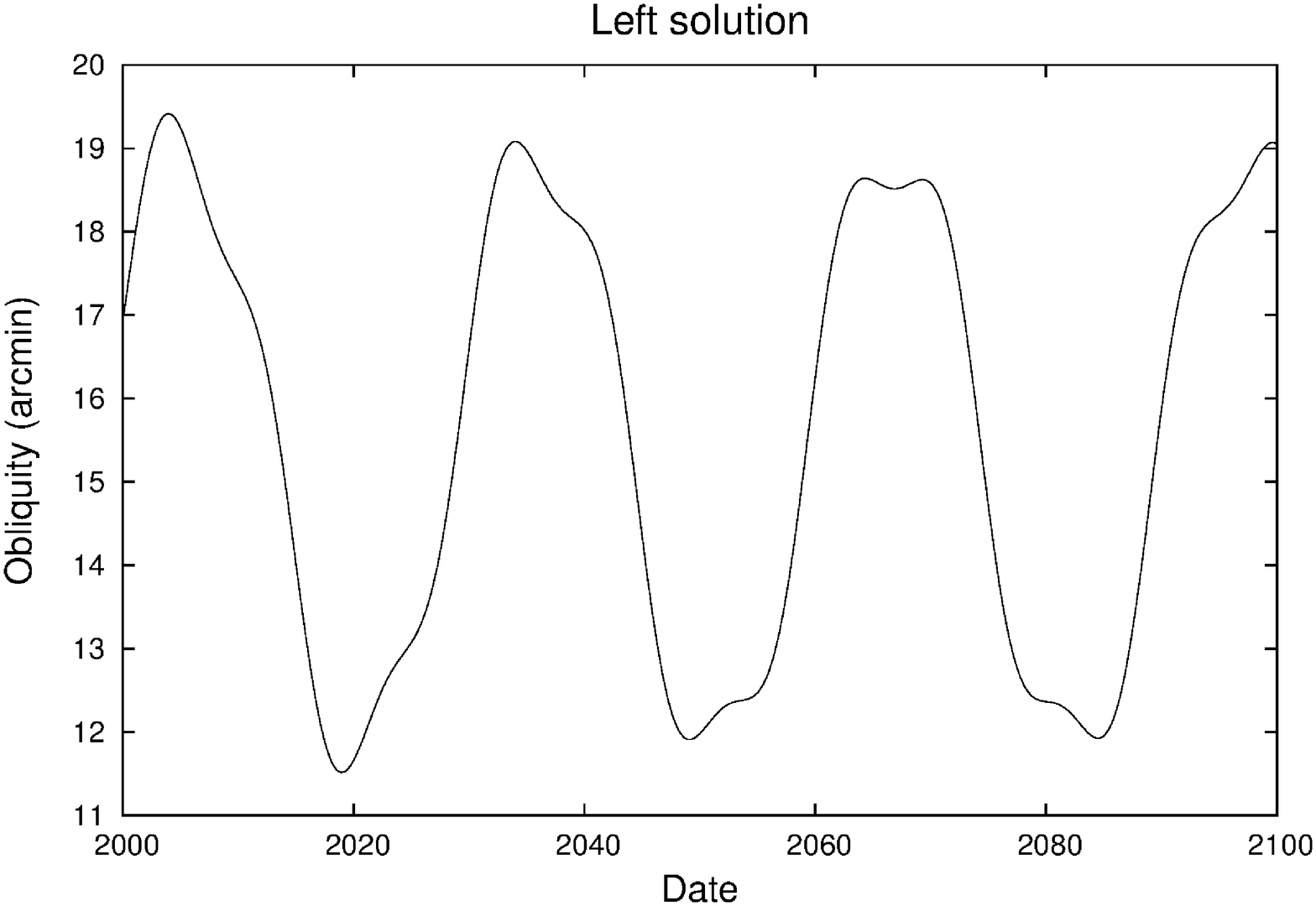} & \includegraphics[width=0.45\textwidth]{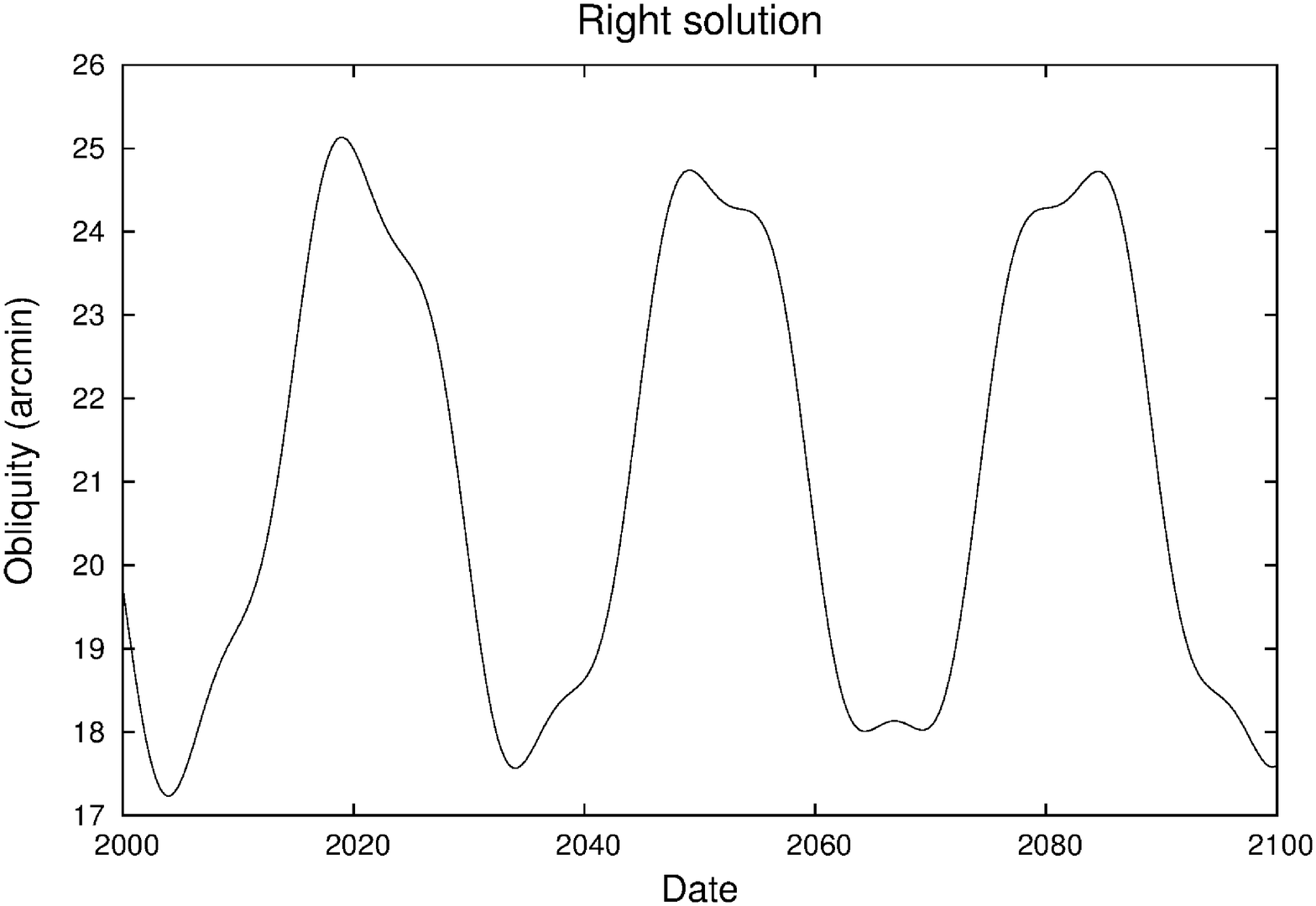}
\end{tabular}
\caption{Left ($T_4 = 29.3$ years) and right ($T_4 = 29.572$ years) solutions over 100 years.\label{fig:leftright100}}
\end{figure}

\begin{figure}[ht]
\centering
\begin{tabular}{cc}
  \includegraphics[width=0.45\textwidth]{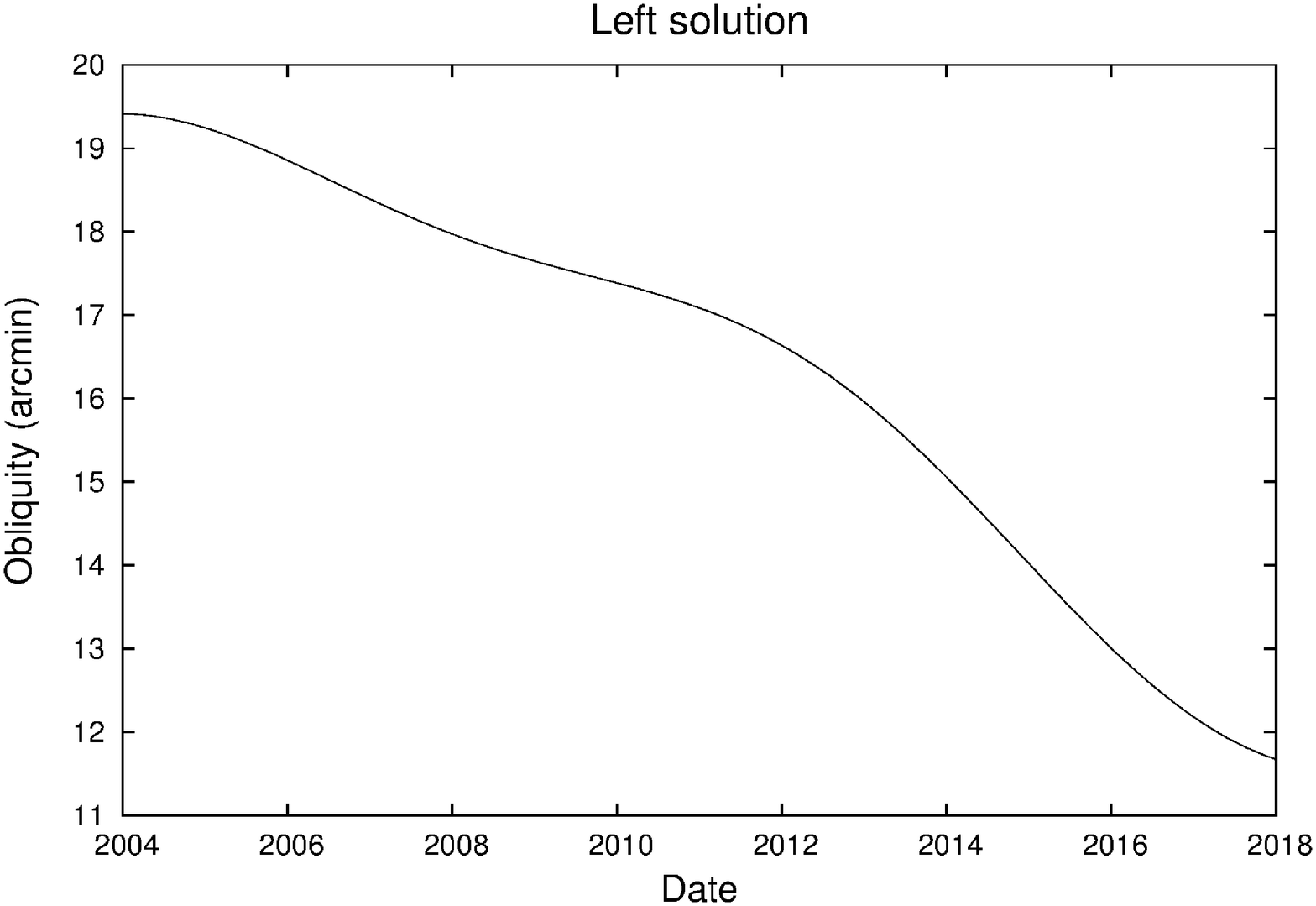} & \includegraphics[width=0.45\textwidth]{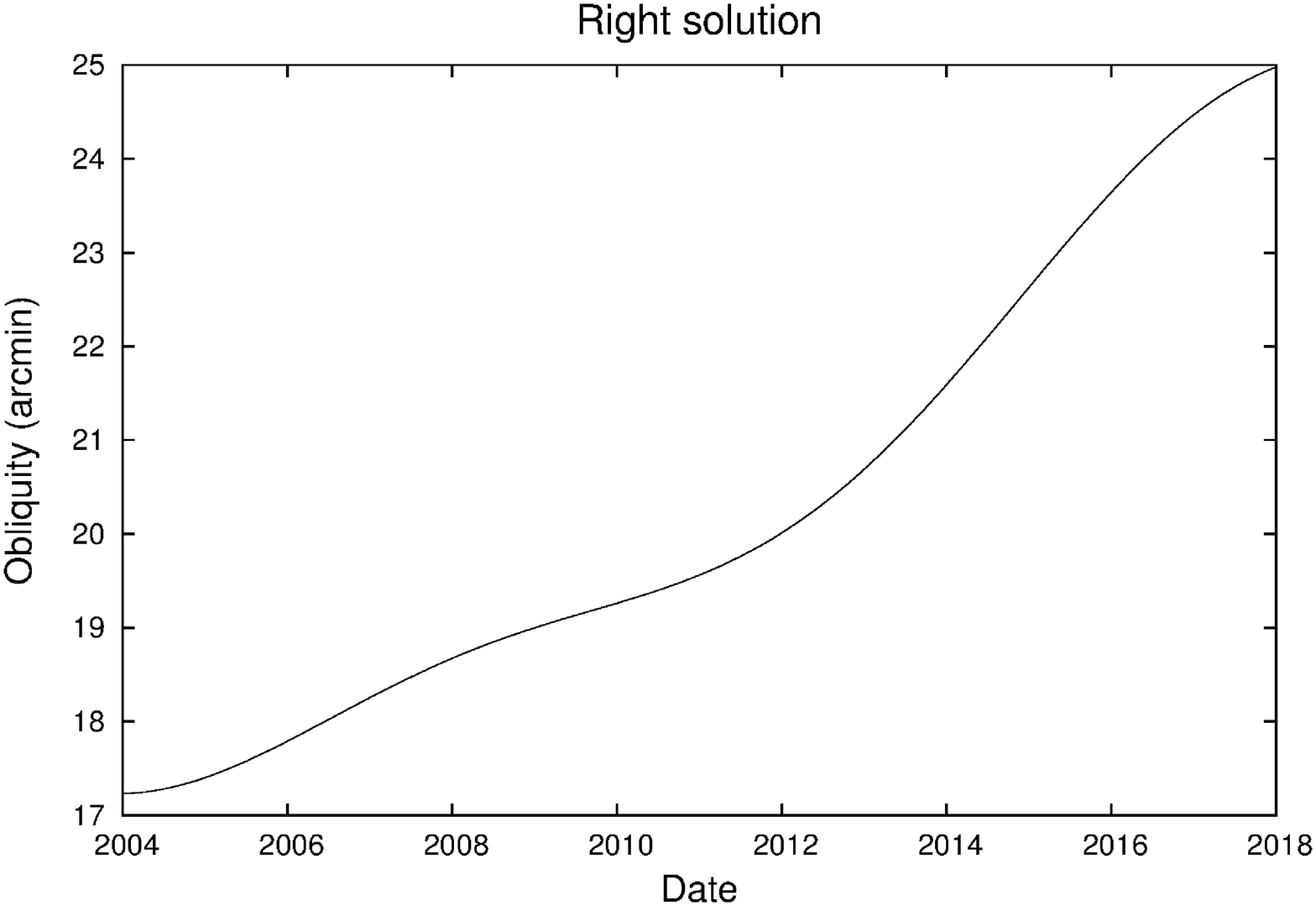}
 \end{tabular}
\caption{Left and right solutions over the {\em Cassini} mission.\label{fig:leftrightcassini}}
\end{figure}

For left solutions we have $T_4=29.3\pm0.03$ years, and for right solutions $T_4=29.572\substack{+0.019 \\ -0.015}$ years. These solutions
are displayed in the Fig.\ref{fig:leftright100} over 100 years and in the Fig.\ref{fig:leftrightcassini} over the duration of the {\em Cassini} mission.

\placefigure{fig:leftright100}

These two solutions coincide at the date of observation, but have actually very different behaviors. They have respectively a mean obliquity
of $\approx15$ and $\approx21$ arcmin because the right solution is closer to the exact resonance than the left one, and they have the same 
period of main oscillation, i.e. $30.74475$ years (Eq.\ref{eq:phi1}) but with opposite phases.

\placefigure{fig:leftrightcassini}

\par As a consequence of these opposite phases, the left solution has a negative slope over the {\em Cassini} mission, while the right one has a positive one 
(see Tab.~\ref{tab:slopes}). A change in obliquity of roughly 7 arcmin (0.12$^\circ$e) over the course of the {\em Cassini} mission is twice 
the formal uncertainty quoted by \citet{mi2012} and should be detectable with careful analysis \citep{bsk2013}.

\placetable{tab:slopes}

\begin{table}[ht]
 \centering
 \caption[Mean predicted time-derivatives of the obliquity of Titan.]{Mean slopes of the obliquity of Titan. The unit is arcsec/y.\label{tab:slopes}}
 \begin{tabular}{c|cc}
  & Left         & Right \\
  & $T_4=29.3$ y & $T_4=29.572$ y \\
 \hline
 $[2004;2012]$ & $-20.8449$ & $20.8449$ \\
 $[2012;2018]$ & $-49.6699$ & $49.6699$ \\
\hline 
 \end{tabular}
\end{table}

\clearpage

\subsection{The obliquity of the core}

\par We also derived the mean obliquity of the core $<K^c>$, from a frequency analysis of the resulting trajectories. It is close to $9.2$ arcmin, and the instantaneous 
core obliquity $K^c(t)$ can vary between $7$ and $10.3$ arcmin over the next 100 years. We also checked that the period $T_3$ of the free oscillations is slightly bigger 
than suggested by the analytical formula (\ref{eq:sigma3}), the error being smaller than $1\%$.

\section{Conclusion}

\par The goal of this study was to investigate the constraint that the rotation of Titan could provide on its interior. Supporting the suggestion originally by \citep{bvyk2011},
we find that between 10 and 13$\%$ of our realistic Titans fall into a resonance with the annual forcing, raising the obliquity of the shell. These Titans have a
130 to 140 km mean thickness shell overlying a $\approx$250 thick ocean, and include shell thickness variations (bottom loading) that are from $80\%$ to $92\%$ compensated, consistent with the gravity and topography constraints.
A better determination of the gravity field would help to refine these numbers.

\par The quasi-resonant behavior results in two solutions to explain the observed obliquity of Titan, that could be discriminated by measuring the time derivative of the 
obliquity. A detection by {\em Cassini} of a time-variable obliquity would thus provide strong evidence for the analysis presented here.

%

\section*{Acknowledgments}

\par Beno\^it Noyelles is an F.R.S.- FNRS postdoctoral research fellow. FN acknowledges support from the Cassini Participating Scientist Program.

\appendix

\section{Derivation of the gravitational torque of the shell on the crust\label{sec:szetoxu}}

We here aim at deriving the gravitational torque of the shell on the crust as \citep{sx1997} did. We first express the vector $\vec{r}$ pointing 
to the position of a mass element of the core, in the reference frames of the shell and of the core:

\begin{eqnarray}
	\vec{r} & = & X\vv{f_1^s}+Y\vv{f_2^s}+Z\vv{f_3^s} \nonumber \\
	& = & x\vv{f_1^c}+y\vv{f_2^c}+z\vv{f_3^c} \nonumber
\end{eqnarray}	
with

\begin{equation}
	\label{eq:l1n3}
	\left(\begin{array}{c}
	X \\
	Y \\
	Z \end{array}\right) = \left(\begin{array}{ccc}
	l_1 & l_2 & l_3 \\
	m_1 & m_2 & m_3 \\
	n_1 & n_2 & n_3 \end{array}\right)\left(\begin{array}{c}
	x \\
	y \\
	z \end{array}\right)
\end{equation}
and 

\begin{equation}
	\label{eq:defl1n3}
	\left(\begin{array}{ccc}
	l_1 & l_2 & l_3 \\
	m_1 & m_2 & m_3 \\
	n_1 & n_2 & n_3 \end{array}\right) = R_3(-\theta^s)R_1(-\epsilon^s)R_3(h^c-h^s)R_1(\epsilon^c)R_3(\theta^c),
\end{equation}
$(X,Y,Z)$ and $(x,y,z)$ being the cartesian coordinates of the mass element expressed in the two reference frames, of the shell and of the core respectively.

\par From

\begin{eqnarray}
	r^2 & = & X^2+Y^2+Z^2, \nonumber \\
	\cos\psi & = & Z/\sqrt{X^2+Y^2+Z^2}, \nonumber \\
	\sin^2\psi\cos 2\phi & = & (X^2-Y^2)/(X^2+Y^2+Z^2), \nonumber \\
	P_2(\cos\psi) & = & (2Z^2-X^2-Y^2)/(2(X^2+Y^2+Z^2)), \nonumber \\
	P_2^2(\cos\psi)\cos 2\phi & = & 3(X^2-Y^2)/(X^2+Y^2+Z^2), \nonumber
\end{eqnarray}
we get

\begin{equation}
	\label{eq:potPhi}
	\Phi = \beta(2Z^2-X^2-Y^2)/2+3\gamma(X^2-Y^2).
\end{equation}

\par After expression of the potential $\Phi$ in the reference frame of the core following Eq.(\ref{eq:l1n3}), derivation to get the gradient and the cross
product, and elimination of the crossed terms resulting in a null integral because the core is triaxial, we get from the Eq.(\ref{eq:torquesx}):

\begin{eqnarray}
	\label{eq:torquesx2}
	\vv{\Gamma^c_{sh}} & = & \beta \left[\begin{array}{c}
	(-2n_2n_3+m_2m_3+l_2l_3)\iiint_{core}\rho_c\left(z^2-y^2\right)dxdydz \vv{f_1^c} \\
	(2n_1n_3-m_1m_3-l_1l_3)\iiint_{core}\rho_c\left(z^2-x^2\right)dxdydz \vv{f_2^c} \\
	(-2n_1n_2+m_1m_2+l_1l_2)\iiint_{core}\rho_c\left(y^2-x^2\right)dxdydz \vv{f_3^c} \end{array}\right] \nonumber \\
	& + & 6\gamma\left[\begin{array}{c}
	(m_2m_3-l_2l_3)\iiint_{core}\rho_c\left(z^2-y^2\right)dxdydz \vv{f_1^c} \\
	(l_1l_3-m_1m_3)\iiint_{core}\rho_c\left(z^2-x^2\right)dxdydz \vv{f_2^c} \\
	(m_1m_2-l_1l_2)\iiint_{core}\rho_c\left(y^2-x^2\right)dxdydz \vv{f_3^c} \end{array}\right]
\end{eqnarray}	

\par Since the matrix of the transformation from $\left(\vv{f_1^s},\vv{f_2^s},\vv{f_3^s}\right)$ to $\left(\vv{f_1^c},\vv{f_2^c},\vv{f_3^c}\right)$ is 
orthogonal as a product of orthogonal matrices, we have 

\begin{eqnarray}
	l_1l_2+m_1m_2+n_1n_2 & = & 0, \nonumber \\
	l_1l_3+m_1m_3+n_1n_3 & = & 0, \nonumber \\
	l_2l_3+m_2m_3+n_2n_3 & = & 0. \nonumber
\end{eqnarray}
And from 

\begin{eqnarray}
	\iiint_{core}\rho_c\left(z^2-y^2\right)dxdydz & = & B^c-C^c, \nonumber \\
	\iiint_{core}\rho_c\left(z^2-x^2\right)dxdydz & = & A^c-C^c, \nonumber \\
	\iiint_{core}\rho_c\left(y^2-x^2\right)dxdydz & = & A^c-B^c, \nonumber
\end{eqnarray}
we finally obtain

\begin{eqnarray}
	\label{eq:torquesx3b}
	\vv{\Gamma^c_{sh}} & = & 3\beta \left[\begin{array}{c}
	n_2n_3(C^c-B^c) \vv{f_1^c} \\
	-n_1n_3(C^c-A^c) \vv{f_2^c} \\
	n_1n_2(B^c-A^c) \vv{f_3^c} \end{array}\right] + 6\gamma\left[\begin{array}{c}
	(l_2l_3-m_2m_3)(C^c-B^c) \vv{f_1^c} \\
	-(l_1l_3-m_1m_3)(C^c-A^c) \vv{f_2^c} \\
	(l_1l_2-m_1m_2)(B^c-A^c) \vv{f_3^c} \end{array}\right],
\end{eqnarray}	
this formula being consistent with the Eq.7 of \citep{sx1997}.

\section{The NAFF algorithm\label{sec:naff}}

\par  The frequency analysis algorithm that we use is based on Laskar's original idea, named NAFF as Numerical Analysis of the 
Fundamental Frequencies (see for instance \citep{l1993} for the method, and \citep{l2005} for the 
convergence proofs). It aims at identifying the coefficients $a_k$ and $\omega_k$ of a complex signal $f(t)$ obtained numerically 
over a finite time span $[-T;T]$  and verifying

\begin{equation}
\label{equ:naff}
f(t) \approx \sum_{k=1}^na_k\exp(\imath\omega_kt),
\end{equation}
where $\omega_k$ are real frequencies and $a_k$ complex coefficients. If the signal $f(t)$ is real, its frequency spectrum is 
symmetric and the complex amplitudes associated with the frequencies $\omega_k$ and $-\omega_k$ are complex conjugates. The 
frequencies and amplitudes associated are found with an iterative scheme. To determine the first frequency $\omega_1$, one searches 
for the maximum of the amplitude of 

\begin{equation}
\label{equ:philas}
\phi(\omega)=<f(t),\exp(\imath\omega t)>,
\end{equation}
where the scalar product $<f(t),g(t)>$ is defined by

\begin{equation}
\label{equ:prodscal}
<f(t),g(t)>=\frac{1}{2T}\int_{-T}^T f(t)g(t)^*\chi(t) dt,
\end{equation}
$g(t)^*$ being the complex conjugate of $g(t)$. $\chi(t)$ is a weight function alike a Hann or a Hamming window, i.e. a 
positive function verifying

\begin{equation}
\label{equ:poids}
\frac{1}{2T}\int_{-T}^T \chi(t) dt=1.
\end{equation}
Using such a window can help the determination in reducing the amplitude of secondary minima in the transform (\ref{equ:prodscal}). 
Its use is optional.
\par Once the first periodic term $\exp(\imath\omega_1t)$ is found, its complex amplitude $a_1$ is obtained by orthogonal 
projection, and the process is started again on the remainder $f_1(t)=f(t)-a_1\exp(\imath\omega_1t)$. The algorithm stops when 
two detected frequencies are too close to each other, what alters their determinations, or when the number of detected terms reaches 
a limit set by the user. This algorithm is very efficient, except when two frequencies are too close to each other. In that case, the 
algorithm is not confident in its accuracy and stops. When the difference between two frequencies is larger than twice the frequency 
associated with the length of the total time interval, the determination of each fundamental frequency is not perturbed by the other 
ones. Although the iterative method suggested by \citep{c1998} allows to reduce this distance, some troubles may 
remain. In our specific case, the Titans affected by the annual resonance present a frequency of free oscillations that is very close
to the forcing frequency of the Sun, the period associated being 29.46 years. For these Titans, distinguishing these two oscillations
is challenging.

\section{Analytical expression of the longitudinal librations\label{sec:analibr}}

\par This calculation is not original and can be found for instance in \citep{bv2010}. We here write it in a way pretty similar as in \citep{vbt2013}. 
Our physical model is different because we do not consider elastic effects.

\par Since we are here only interested in the longitudinal librations, we can a priori assume that the obliquities and polar motions of the core and the shell are negligible.
This yields $\omega_1^c=\omega_2^c=\omega_1^s=\omega_2^s=0$ and $\epsilon^c=\epsilon^s=0$, i.e. $\xi^c=\eta^c=\xi^s=\eta^s=0$. So, only the Eq. \ref{eq:dotpc}, 
\ref{eq:dotomega3c}, \ref{eq:dotps} and \ref{eq:dotomega3s} still hold. After expansion of the orbital ephemerides up to the first order in eccentricity $e_6$ and expression of the 
resonant arguments $\gamma^c = p^c-\lambda_6$ and $\gamma^s = p^s-\lambda_6$, we get:

\begin{eqnarray}
  C^s\ddot{\gamma^s}+K_1\gamma^s+K_2\gamma^c & = & 4eK_3\sin\mathcal{M}_6, \label{eq:librashell} \\
  C^c\ddot{\gamma^c}+K_2\gamma^s+K_4\gamma^c & = & 4eK_5\sin\mathcal{M}_6, \label{eq:libracore} 
\end{eqnarray}
with

\begin{eqnarray}
  K_1 & = & 3n_6^2\left(B^s-A^s+B_t^o-A_t^o\right)+12\gamma^o\left(B^c-A^c+B_b^o-A_b^o\right), \label{eq:K1} \\
  K_2 & = & -12\gamma^o\left(B^c-A^c+B_b^o-A_b^o\right), \label{eq:K2} \\
  K_3 & = & \frac{3}{2}n_6^2\left(B^s-A^s+B_t^o-A_t^o\right), \label{eq:K3} \\
  K_4 & = & \left(3n_6^2+12\gamma^o\right)\left(B^c-A^c+B_b^o-A_b^o\right), \label{eq:K4} \\
  K_5 & = & \frac{3}{2}n_6^2\left(B^c-A^c+B_b^o-A_b^o\right), \label{eq:K5}
\end{eqnarray}
$\mathcal{M}_6$ being the mean anomaly of Titan. The amplitudes of the forced librations at orbital period $g^s$ and $g^c$ can be written as

\begin{eqnarray}
  g^s & = & \frac{4e\left(n_6^2K_3C^c+K_2K_5-K_3K_4\right)}{C^cC^s\left(n_6^2-\sigma_1^2\right)\left(n_6^2-\sigma_2^2\right)}, \label{eq:gs} \\
  g^c & = & \frac{4e\left(n_6^2K_5C^s+K_2K_3-K_1K_5\right)}{C^cC^s\left(n_6^2-\sigma_1^2\right)\left(n_6^2-\sigma_2^2\right)}, \label{eq:gc}
\end{eqnarray}

where $\sigma_1$ and $\sigma_2$ are the frequencies of the free longitudinal oscillations:

\begin{equation}
  \label{eq:sigma12}
  \sigma_{1,2} = \frac{K_1C^c+K_4C^s\pm\sqrt{4(K_2^2-K_1K_4)C^cC^s+\left(K_1C^c+K_4C^s\right)^2}}{2C^cC^s}.
\end{equation}

\section{Notations used in this paper}

\placetable{tab:notations}

\begin{deluxetable}{lr}
	\tablecolumns{2}
	\tablewidth{0pt}
	\tablecaption{Main notations used in this study.\label{tab:notations}}
	\startdata
 \hline
  a,b,c,R                                                                                      & External radii and mean radius of Titan \\
  $h^c$,$\epsilon^c$,$\theta^c$                                                                & Euler angles orienting the principal axes of the core \\
  $h^s$,$\epsilon^s$,$\theta^s$                                                                & Euler angles orienting the principal axes of the shell \\
  $\xi^c$,$\eta^c$,$p^c$                                                                       & Non-singular Euler coordinates for the core \\
  $\xi^s$,$\eta^s$,$p^s$                                                                       & Non-singular Euler coordinates for the shell \\
  $\vec{G^c}$,$\vec{G^s}$                                                                      & Angular momentum of the shell and the core \\
  $(\vv{f_1^c},\vv{f_2^c},\vv{f_3^c})$                                                         & Reference frame of the principal axes of inertia of the core  \\
  $(\vv{f_1^s},\vv{f_2^s},\vv{f_3^s})$                                                         & Reference frame of the principal axes of inertia of the shell \\
  $A^c$,$B^c$,$C^c$,$A^s$,$B^s$,$C^s$                                                          & Principal moments of inertia of the core and the shell \\
  $A^o_t$,$B^o_t$,$C^o_t$,$A^o_b$,$B^o_b$,$C^o_b$                                              & Principal moments of inertia of the top and bottom oceans \\
  $\vv{r_{\saturn}^s}=x_{\saturn}^c\vv{f_1^c}+y_{\saturn}^c\vv{f_2^c}+z_{\saturn}^c\vv{f_3^c}$ & Vector Titan-Saturn in the frame of the core \\
  $\vv{r_{\saturn}^s}=x_{\saturn}^s\vv{f_1^s}+y_{\saturn}^s\vv{f_2^s}+z_{\saturn}^s\vv{f_3^s}$ & Vector Titan-Saturn in the frame of the shell \\
  $M_{\saturn}$,$M_6$                                                                          & Masses of Saturn and of Titan \\
  $n_6$,$\mathcal{M}_6$,$\lambda_6$                                                            & Mean motion, anomaly and longitude of Titan \\
  $e_6$,$I_6$                                                                                  & Eccentricity and inclination of Titan \\
  $\varpi_6$,$\ascnode_6$                                                                      & Longitudes of the pericentre and the ascending node of Titan \\
  $\rho_s$,$\rho_o$,$\rho_c$,$\rho_6$                                                          & Densities of the shell, the ocean, the core, and Titan \\
  $a_o$,$b_o$,$c_o$,$a_c$,$b_c$,$c_c$                                                          & Radii of the shell-ocean boundary and of the core \\
  $f_1$,$\kappa_1$                                                                             & Flattening and equatorial ellipticity of the shell-ocean boundary \\
  $f_2$,$\kappa_2$                                                                             & Flattening and equatorial ellipticity of Titan \\
  $\psi$,$\phi$                                                                                & Colatitude and east longitude of a mass element \\
  $\mathcal{G}$                                                                                & Gravitational constant \\ 
  $\vv{\Gamma_{\saturn}^{s,o}}$                                                                & Gravitational torque of Saturn on the shell+top ocean \\
  $\vv{\Gamma_{\saturn}^{c,o}}$                                                                & Gravitational torque of Saturn on the core+bottom ocean \\
  $\vv{\Gamma_{co}^{s,o}}$                                                                     & Gravitational torque of the core on the shell+top ocean \\
  $\vv{\Gamma_{sh}^{c,o}}$                                                                     & Gravitational torque of the shell on the core+bottom ocean \\
  $\psi^c$,$\psi^s$                                                                            & Tidal librations of the core and the shell \\
  $\gamma^c$,$\gamma^s$                                                                        & Physical librations of the core and the shell \\
  $g^c$,$g^s$                                                                                  & Amplitude of the diurnal component of the physical librations \\
  $K^c$,$K^s$                                                                                  & Obliquities of the core and the shell \\
  $Q_1^c$,$Q_2^c$,$Q_1^s$,$Q_2^s$                                                              & Components of the polar motions of the core and the shell \\
  $d_o$,$d_s$                                                                                  & Mean thicknesses of the shell and the ocean \\
  $h_t$,$h_b$                                                                                  & Topographic top and bottom anomalies \\
  $f$                                                                                          & Isostatic compensation factor \\
  $\sigma_1$,$\sigma_2$,$T_1$,$T_2$                                                            & Frequencies and periods of the free librations in longitude \\
  $\sigma_3$,$\sigma_4$,$T_3$,$T_4$                                                            & Frequencies and periods of the free librations of the obliquity \\
  \hline
  \enddata
\end{deluxetable}

\clearpage

\listoftables

\clearpage

\clearpage

\listoffigures

\end{document}